\documentclass[journal,comsoc]{IEEEtran} 
\normalsize

\usepackage[utf8]{inputenc} 
\usepackage[T1]{fontenc}
\usepackage{url,comment}
\usepackage{ifthen}
\usepackage{cite}
\usepackage{graphicx}
\usepackage{amsmath,amssymb,amsthm}
\usepackage{cite,soul}
\usepackage{floatrow}
\usepackage{xcolor}
\usepackage{enumitem}

\usepackage{multirow,rotating}
\usepackage{diagbox}
\usepackage{threeparttable}

\DeclareMathOperator{\expD}{Exp}
\DeclareMathOperator{\sexpD}{S-Exp}

\newtheorem{theorem}{Theorem}
\newtheorem{lemma}{Lemma}

\newtheorem{conjecture}{Conjecture}

\newtheorem{remark}{Remark}
\usepackage{booktabs}
\usepackage[colorlinks,pdfstartview=FitH,citecolor=blue,linkcolor=blue,urlcolor=blue]{hyperref}
\usepackage{tikz}
\usepackage[capitalize]{cleveref}

\newcommand{\prob}[1]{{\mathbb P}}

\usepackage[normalem]{ulem}
\usepackage{xcolor}
\makeatletter
\def\squiggly{\bgroup \markoverwith{\textcolor[rgb]{1,0,0}{\lower3.5\p@\hbox{\sixly \char58}}}\ULon}
\makeatother

\ifCLASSINFOpdf
\else
\fi

\begin{document}
%
\title{Distributed  Storage  Allocations\\ for  Optimal Service Rates}
%
%
%
\author{Pei Peng, Moslem Noori, and Emina Soljanin,~\IEEEmembership{Fellow,~IEEE}
\thanks{P.~Peng and E.~Soljanin  are with the ECE Department at Rutgers, The State University of New Jersey, Piscataway, NJ 08854, USA, e-mail: pei.peng@rutgers.edu, (see https://www.ece.rutgers.edu/emina-soljanin). M.~Noori is with 1 QB Information Technologies (1QBit), Vancouver, BC, Canada, e-mail: moslem.noori@gmail.com.\\
Some parts of Sec.~\ref{sec:small}, \ref{sec:large_exp} and \ref{sec:large_exp} of this paper appeared in the Proc.\ of 2016 IEEE Internat.\ Symp.\ on Information Theory (ISIT) \cite{noori2016storage} and 2018 56th Annual Allerton Conf.\ on Communication, Control, and Computing \cite{peng2018distributed}.\\ Part  of  this  research  is  based  upon  work  supported  by  the  National  Science  Foundation  under  Grant  No.  CIF-1717314.}
}%
%
%

\markboth{IEEE Transactions on Communications, to appear in  2021}%
{Submitted paper}
%



\maketitle
\begin{abstract}

 Distributed systems operate under storage access and download service uncertainty. We consider two access models. In one, a user can access each storage node with a fixed probability, and in the other, a user can access any fixed-size subset of nodes. We consider two download service models. In the first (small file) model, the time to transmit file data is negligible compared to the overall average download time. In the second (large file) model, the download time scales with the amount of downloaded data. The performance metric is the system’s service rate. For a fixed redundancy level, the systems’ service rate depends on the allocation of coded chunks over the storage nodes. Since finding the general optimal allocation is prohibitively hard, we consider quasi-uniform allocations, where coded content is equally spread among a subset of nodes. The question we address asks what the size of this subset (spreading) should be. We show that concentrating the coded content to a minimum-size subset is universally optimal for the small file model. However, for the large file model, the optimal spreading depends on the system parameters. These conclusions hold for both access models.
\end{abstract}

\begin{IEEEkeywords}
Distributed storage systems, service rate, optimal allocations, erasure coding, redundancy. 
\end{IEEEkeywords}

%

\IEEEpeerreviewmaketitle

\section{Introduction}

Distributed storage systems (DSSs) are a vital part of computing and content providing environments, such as cloud data centers and edge systems. Their purpose is to ensure reliable storage and quick access of data by end-users or computing processes. Today, both goals are commonly addressed by storing data redundantly. The DSS performance must be robust to various forms of uncertainty.
Most of the current work addresses internal uncertainty (e.g., straggling) in operations of the system itself (see, e.g., \cite{joshi2012coding, kadhe15availability, aktas2019straggler,aktas2021download}). This paper considers the uncertainty in both network accessibility and download services, which are common in edge computing \cite{Edge:YadgarKAS19}.

 We address the following network access uncertainties, as considered in  \cite{leong2011distributed} and the follow-up work. 1) Users may only be able to access a random subset of nodes. Such users can retrieve a file if the storage content of the accessed nodes suffices for file decoding. 2) Even when users have access to all nodes, a node may not respond. Such users can retrieve the file if the storage content of the responding nodes suffices for file decoding. We adopt a storage model
 of \cite{leong2011distributed}, where files are split into  chunks, and redundancy is introduced at some fixed level, determined by the storage budget that the DSS has for the file. The total storage is the only constraint. There is no limit on how many chunks a particular node can store.

We consider two download service models. In the first download model, the time it takes to transmit file data is negligible compared with the overall average download time. Thus the download time is a random variable that only depends on the storage system parameters. 
We refer to this model as the {\it small file} scenario.
In the second download model, there is randomness associated with both the file transmission and inherent system’s operations. 
Thus the download time scales with the amount of data being downloaded.
We refer to this model as the {\it large file} scenario. We adopt two common service and scaling models used in the literature. For more detail and other models, see \cite{peng2020diversity,peng2020diversityTIT} and references therein.

Two important DSS performance measures have been considered in the literature \cite{leong2011distributed,leong2012distributed,Sardari_Allocation_2010,hong2014asymptotic, noori2015allocation,noori2016storage, peng2018distributed,wu2020optimal}. One is the {\it probability of successful data recovery} and the other is the {\it average service rate}.  Finding these quantities has been challenging, and the optimal allocations are known only in some special cases. Some versions of this problem are related to a long-standing conjecture by Erd\H{o}s on the maximum number of edges in a uniform hypergraph \cite{matching:AlonFHRRS12}.
In general, both measures are of interest and should be simultaneously taken into account. Increasing the chance of successful download, while desirable, should not come at the cost of intolerable delivery delay. Moreover, in practice, we may want to partially sacrifice a successful but tardy data delivery to some users to ensure that other users, that can receive the data, are served fast. 
This paper focuses on the service rate of a DSS. Service rate is an emerging increasingly important performance measure, which addresses stability in distributed systems with redundancy under uncertainty in service request arrivals \cite{ service:aktas2020service,service:KazemiKSS21g,service:KazemiKSS20g,service:KazemiKSS20c,service:AndersonJJ18,service:AktasAJ17,service:RaaijmakersB20}.

For a fixed redundancy level, the system's service rate is determined by the allocation of coded chunks over the storage nodes. We consider quasi-uniform storage allocations, where coded content is uniformly spread among a subset of nodes, and ask what the size of this subset (spreading) should be to maximize the expected download service rate.
We consider two service time models: scaled exponential service time and shifted exponential service time. These and other models are considered in the context of the server-dependent scaling and data-dependent scaling in \cite{peng2020diversity, peng2020diversityTIT}.  We show that concentrating the coded content to a minimum-size subset is universally optimal for the small file model. However, for the large file model, the optimal spreading depends on the system parameters.  These  conclusions  hold  for  both  access  models.

The paper is organized as follows: In Sec.~\ref{sys:model}, we present the system architecture and the models for the service time and rate. In Sec.~\ref{sec:contributions}, we state the problem and summarize the contributions of this paper. In Sec.~\ref{sec:small}, \ref{sec:large_exp}, and \ref{sec:large_shift}, we characterize the DSS service rate and determine the optimal allocation for three common service time distributions and two different access models.  
Conclusions are given in Sec.~\ref{sec:con}.

\section{System Model and Problem Formulation}
\label{sys:model}

\subsection{Storage Model}
\label{sec:StorageModel}
A file consisting of $k$ blocks is redundantly stored over a DSS with $N$ storage nodes. The file is encoded by a maximum distance separable (MDS) code into $mk$ ($m \in \mathbb{N}$) encoded blocks so that any $k$ of them are sufficient to recover the file. The $mk$ encoded blocks are partitioned into $N$ subsets $\mathcal{S}_i$'s for $i  \in \{1,\dots, N\}$ where $| \mathcal{S}_i| = s_i$, and thus $\sum_{i = 1}^N s_i = mk$. We refer to such partitioning as {\it allocation}. The $s_i$ blocks in $\mathcal{S}_i$ are stored at the storage node $i$. Note that $0 \leq s_i \leq k$ since storing more than $k$ blocks on a node is unnecessary.

 \begin{figure*}[hbt]
    \centering
    \includegraphics[width=0.68\textwidth]{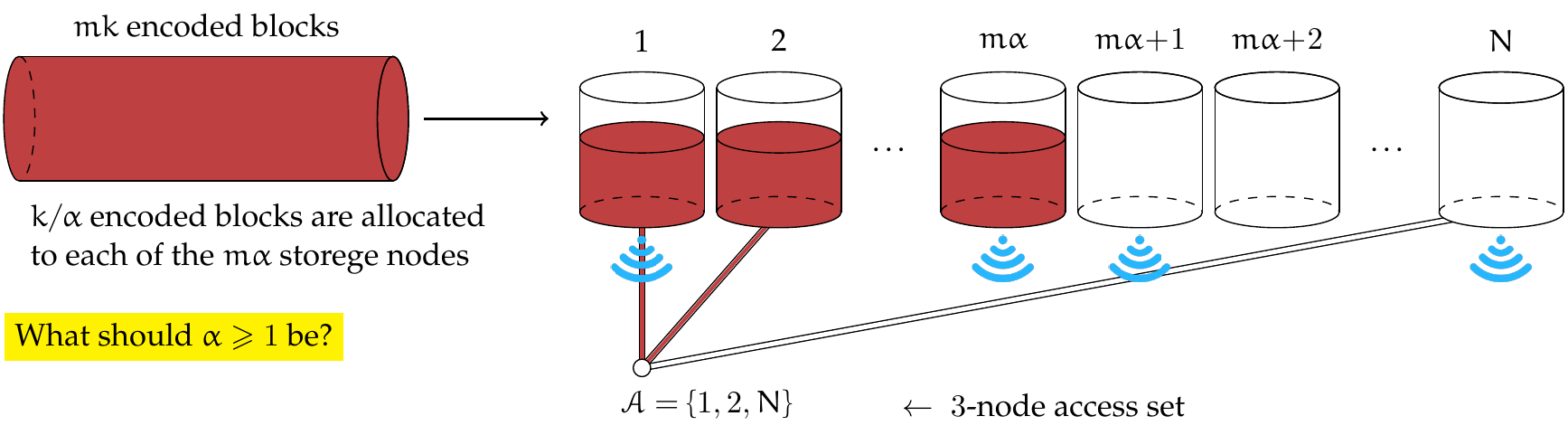}
    \caption{A DSS with $N$ nodes with quasi-uniform allocation. Each node stores either $k/\alpha$ or $0$ data blocks of interest to some users, and thus only $\varphi=m\alpha$ nodes contain data blocks. The WiFi sign indicates that the node is able to serve the user. Note that that is independent of whether or not the node has been accessed or has the data. Here, three nodes are successfully accessed, but only two of them have (coded) data blocks. One of the accessed nodes has data blocks but is not able to serve the user.}
    \label{fig:DSS}
\end{figure*}

Optimizing general storage allocations is computationally difficult, see \cite{leong2012distributed}. Thus we focus on quasi-uniform allocations \cite{Sardari_Allocation_2010}, where a node can either store a constant number of blocks $k/\alpha$ ($\alpha \in \mathbb{N}$) or no blocks at all.
We will refer to such allocation as $\alpha$ quasi-uniform allocation. Fig.~\ref{fig:DSS} depicts an example of $\alpha$ quasi-uniform allocation on $N$ nodes.
We refer to a quasi-uniform allocation where $\alpha = 1$ as \emph{the minimal spreading} allocation \cite{leong2012distributed}. Note that for the minimal spreading allocation,  the $k$ file blocks are simply replicated over some $m$ storage nodes. Similarly, an allocation with $\alpha = N/m$ is referred to as a \emph{the maximal spreading} allocation since the file chunks are spread over all $N$ nodes in the system.

\subsection{Data Access and Delivery Models\label{sec:AccessModel}}
\noindent\ul{Fixed-size Access:}
In this model, the download request is forwarded to a random $r$-node subset of the $N$ storage nodes. 
Therefore, since the data is MDS encoded, the access to a given $r$-subset $\mathcal{A}$ results in the successful recovery of the data iff the nodes in $\mathcal{A}$ jointly contain at least $k$ coded blocks: 
\begin{equation}\label{eq:RecCond}
\sum_{i \in \mathcal{A}} s_i \ge k.
\end{equation}
\ul{Probabilistic Access:}
Here, each download request is forwarded to all $N$ nodes. However, a node does not respond with probability $p$. Let $\mathcal{A}$ be the set of nodes that are successfully accessed. Then the condition for data recovery is again (\ref{eq:RecCond}). In this case, $1 \leq \alpha \leq \frac{N}{m}$. In this access model, $|\mathcal{A}|$ is a binomial random variable distributed as $\text{Bin}(N,p)$. 

 Regardless of the access model, for an accessed subset of nodes $\mathcal{A}$, we denote the number of nodes containing data by $\varphi(\mathcal{A})$. 
 Observe that $\varphi(\mathcal{A})\le|\mathcal{A}|$.
 For instance, in Fig.~\ref{fig:DSS}, three nodes 
($|\mathcal{A}| = \text{3}$) are accessed while only $\varphi(\mathcal{A}) = 2$ of them have data. 
We assume a request is simultaneously served by all nodes in the accessed set $\mathcal{A}$, where each node takes some i.i.d.\ random time to deliver its data blocks. In the fixed-size access model, $|{\mathcal A}| = r$, while in the probabilistic access model, $|\mathcal{A}|$ is a Binomial random variable between 1 and $N$. The file can be reconstructed when the accessed nodes jointly deliver $k$ encoded blocks. We assume that a node has to deliver all its blocks for the download to count. 
For an $\alpha$ quasi-uniform allocation, the download request can be served iff $\varphi(\mathcal{A}) \geq \alpha$, and as soon as all blocks are downloaded from any $\alpha$ out of $\varphi(\mathcal{A})$ nodes. Therefore, the average service time for the file, $T_s(\alpha|\varphi(\mathcal{A}))$, is the expected value of the $\alpha$-th order statistics of $\varphi(\mathcal{A})$ waiting times at the storage nodes.

\subsection{Download-time  Models}\label{subsec:downloadtime}
As discussed in the introduction, depending on how the data transmission time compares with the average time the system takes to fulfill the request, we consider the small and large file models. Here, \textit{small} and \textit{large} are informal descriptive terms.
The precise mathematical models are stated next. 
\subsubsection{Small File Model}
When a task is assigned to a storage node, there is some random waiting time before the data transmission starts 
(needed, e.g., for a general handshake and/or acquiring the requested content). We assume the waiting time follows an \ul{exponential distribution}. Compared to the mean of this distribution, the data transmission time is negligible.

The waiting times at nodes are independent random variables, each following an exponential distribution $\expD(\mu)$ with mean $1/\mu$. Thus the average service time $T_s(\alpha|\varphi(\mathcal{A}))=\frac{1}{\mu}(H_{\varphi(\mathcal{A})}-H_{\varphi(\mathcal{A})-\alpha})$,
where $H_{\alpha}=\sum^{\alpha}_{i=1}1/i$ is the $\alpha$-th 
harmonic number \cite{arnold2008first}.
The corresponding service rate achieved by the nodes in $\mathcal{A}$ (with $\varphi(\mathcal{A})>\alpha$ nodes containing data) is
\begin{equation}
\mu_s(\alpha|\varphi(\mathcal{A}))=\frac{1}{T_s(\alpha|\varphi(\mathcal{A}))}=\frac{\mu}{H_{\varphi(\mathcal{A})}-H_{\varphi(\mathcal{A})-\alpha}}.
    \label{eq:ServiceRate_exp}
\end{equation}
It is not hard to see that 
\begin{equation}
 \mu_s(\alpha|\varphi(\mathcal{A}))\le\mu \varphi(\mathcal{A}).
\label{ieq:rate_exp}
\end{equation}
\subsubsection{Large File Model}
Here the download time scales with the number of chunks being downloaded. We consider two distribution/scaling models for the service time: scaled exponential and scaled-shift exponential. 
\\[1ex]
\ul{Scaled Exponential Service Time:} We assume that a node storing the whole file delivers all of its blocks in a random time, exponentially distributed with the mean $1/\mu$, and that 
 a node storing $1/\alpha$ fraction of the file delivers all of its blocks in the random time exponentially distributed with the mean $1/(\alpha\mu)$.
For this model, by applying the order statistics for exponential distribution, we have $T_s(\alpha|\varphi(\mathcal{A}))=\frac{1}{\alpha\mu}(H_{\varphi(\mathcal{A})}-H_{\varphi(\mathcal{A})-\alpha})$,
where $1/(\alpha\mu)$ comes from the service rate scaling discussed above.
The corresponding service rate from set $\mathcal{A}$ is
\begin{equation}
\mu_s(\alpha|\varphi(\mathcal{A}))=\frac{\alpha\mu}{H_{\varphi(\mathcal{A})}-H_{\varphi(\mathcal{A})-\alpha}}.
    \label{eq:SetServiceRate}
\end{equation}
It is not hard to see that 
\begin{equation}
\mu \varphi(\mathcal{A})\ge \mu_s(\alpha|\varphi(\mathcal{A}))\ge \mu (\varphi(\mathcal{A})-\alpha+1).
\label{ieq:rate}
\end{equation}
\ul{Shifted Exponential Service Time:}
Here the data delivery consists of two steps: first, the node takes an exponential random time to process the request; second, it takes a constant time, proportional to its number of the node's stored data blocks, to deliver them to the user. Therefore, the two-step delivery time for a node storing $1/\alpha$ fraction of the file can be modeled by the shifted exponential distribution with rate $\mu$ and the shift parameter $\Delta/\alpha$, denoted by $\sexpD(\Delta/\alpha, \mu)$.
For this model, since shifted exponential distribution is a combination of a constant and an exponential tail, we have $T_s(\alpha|\varphi(\mathcal{A}))=\frac{\Delta}{\alpha}+\frac{1}{\mu}(H_{\varphi(\mathcal{A})}-H_{\varphi(\mathcal{A})-\alpha})$.
The corresponding service rate from set $\mathcal{A}$ is
\begin{equation}
\mu_s(\alpha|\varphi(\mathcal{A}))=\frac{\alpha\mu}{\Delta\mu+\alpha(H_{\varphi(\mathcal{A})}-H_{\varphi(\mathcal{A})-\alpha})}.
    \label{eq:SetServiceRateshift}
\end{equation}
As $\varphi(\mathcal{A}) \in [\alpha,m\alpha]$, it is not hard to see that 
\begin{equation}
\frac{\mu \varphi(\mathcal{A})}{\Delta \mu +\alpha}\ge \mu_s(\alpha|\varphi(\mathcal{A})) \ge \frac{\alpha\mu (\varphi(\mathcal{A})-\alpha+1)}{\Delta \mu(m\alpha-\alpha+1)+\alpha^2}.
\label{ieq:rateshift}
\end{equation}

\subsection{DSS Performance Metrics}\label{subsec:DSSrate}
We consider two key performance metrics: {\it probability of successful data recovery} and {\it average service rate}. 
In general, both measures are of interest and should be simultaneously taken into account. However, as we will see below, they are often maximized by different allocations. In many applications, increasing the chance of successful download is desirable but should not come at the cost of intolerable delivery delay. 
\subsubsection{Probability of File Recovery}
For an $\alpha$ quasi-uniform allocation, data recovery from this subset is successful iff $\varphi(\mathcal{A}) \geq \alpha$. 
The probability of successful file recovery under $\alpha$ quasi-uniform allocation is
\begin{equation}
P_s(\alpha)=
\sum_{\mathcal{A}:\; \sum_{i \in \mathcal{A}} s_i \ge k}
P(\mathcal{A})
\label{eq:prob}
\end{equation}
where $P(\mathcal{A})$ is the probability of accessing $\mathcal{A}$. Note that the sum goes over all sets $\mathcal{A}$ that satisfy the condition \eqref{eq:RecCond}. It follows that $P(\mathcal{A})=\binom {m\alpha}{\varphi(\mathcal{A})}
\binom {N-m\alpha}{r-\varphi(\mathcal{A})}/\binom{N}{r}$ for the fixed-size access and
$P(\mathcal{A})= \binom {m\alpha}{\varphi(\mathcal{A})} (1-p)^{\varphi(\mathcal{A})}p^{m\alpha-\varphi(\mathcal{A})}$ for the probabilistic access.
\subsubsection{Service Rate}
Under an $\alpha$ quasi-uniform allocation, the service rate $\mu_s(\alpha)$, found by averaging over the conditional service rates, is
\begin{equation}
\mu_s(\alpha) = 
\sum_{\mathcal{A}:\; \sum_{i \in \mathcal{A}} s_i \ge k}
P(\mathcal{A})\mu_s(\alpha|\varphi(\mathcal{A}))
\label{eq:service}
\end{equation}
where $\mu_s(\alpha|\varphi(\mathcal{A}))$ is the service rate when the set of accessed nodes is $\mathcal{A}$, given by \eqref{eq:ServiceRate_exp}, \eqref{eq:SetServiceRate} or \eqref{eq:SetServiceRateshift}. When the set $\mathcal{A}$ does not satisfy the condition \eqref{eq:RecCond}, we define $\mu_s(\alpha|\varphi(\mathcal{A}))=0$.


\section{Problem Statement and Contributions
\label{sec:contributions}}
\begin{center}
\begin{small}
    \begin{tabular}{rcl}
       $N$ & -- & number of storage nodes (DSS size) \\
        $m$ & -- & $mk$ is number of encoded blocks\\
       $\mathcal{A}$ & -- & accessed subset of nodes\\
       $\varphi(\mathcal{A})$ & -- & number of nodes in $\mathcal{A}$ containing data \\
       $\alpha$ & -- & $m\alpha$ is the number of nodes with blocks\\
       $P_s(\alpha)$ & -- & probability of successful file recovery\\
        $\mu_s(\alpha)$ & -- & DSS servicce rate\\
        $k$ & -- & number of block in a file \\
        $r$ & -- & number of accessed nodes (fixed-size model)\\
        $p$ & -- & probability of failed access (probabilistic model)\\
    \end{tabular}
\end{small}
\end{center}
System parameters and notations are summarised in the above list.
Our goal is to characterize the DSS service rate $\mu_s(\alpha)$ for the access and service time models defined in Sec.~\ref{sec:AccessModel} and \ref{subsec:downloadtime}. We are in particular interested in finding which $\alpha$ maximizes $\mu_s(\alpha)$. Recall that when $\alpha=1$, we have the minimal spreading allocation, and when $\alpha=N/m$, we have the maximal spreading allocation. When $1<\alpha<N/m$, we have an $\alpha$ quasi-uniform allocation. 

We conclude that the allocation that maximizes the service rate $\mu_s(\alpha)$ depends on the model. For the small file model, the minimal spreading allocation,  i.e. $\alpha = 1$, is always optimal, while for the large file model, this is not the case and it is difficult to determine the optimal allocation. We summarize the regimes where the minimal spreading allocation is optimal and non-optimal for large files in Table~\ref{Table:optimal}.

\begin{table*}[hbt]
		\begin{small}
\floatbox[{\capbeside\thisfloatsetup{capposition={top},capbesidewidth=\textwidth}}]{table}[\FBwidth]
   \caption{\sc{\sc Conditions for the minimal spreading allocation  being optimal/Non-optimal for large files}\label{Table:optimal}}
   \vspace{0.2cm}
   \begin{center}
			\begin{tabular}{@{}lllll@{}}
				\toprule 
				& &   \multicolumn{2}{c@{}}{\sc {Optimality Conditions}}
				\\  \cmidrule{3-4}
				& & {\bf Scaled Exponential}
				& {\bf Shifted Exponential}
				\\ [2mm]
				\multirow{4}{*}{\rotatebox{90}{
						{\sc{Access}}}} & 
				{\bf Fixed-size} &  $r\le \min_{2\le \alpha\le r} \{1+\frac{N-1}{\sqrt[\alpha-1]{\alpha \binom{m\alpha-1}{\alpha-1}}}\}$ &  $r\le \min_{2\le \alpha\le r} \{ 1+\sqrt[\alpha-1]{\frac{\Delta\mu+\alpha}{\alpha(\Delta\mu m+1)\binom{m\alpha-1}{\alpha-1}}}(N-1)\}$\\[2mm] 
				& {\bf Probabilistic}   &$p\ge\max_{\alpha\ge2}\{1-\frac{1}{\sqrt[\alpha-1]{\alpha\binom{m\alpha-1}{\alpha-1}}}\}$ &   $ p\ge\max_{\alpha\ge2}\{1-\sqrt[\alpha-1]{\frac{\Delta\mu+\alpha}{\alpha(\Delta\mu m+1)\binom{m\alpha-1}{\alpha-1}}}\}$\\[2mm]
				\toprule 
				& &   \multicolumn{2}{c@{}}{\sc {Non-optimality Conditions}}
				\\  \cmidrule{3-4}
				\multirow{4}{*}{\rotatebox{90}{
						{\sc{Access}}}} & 
				{\bf Fixed-size} &  $\begin{aligned} r\ge\min_{2\le\alpha\le r}\{\sqrt[\alpha-1]{\frac{m}{m\alpha-\alpha +1}}\\ \cdot(N-\alpha+1)+\alpha-1\}\end{aligned}$ &  $\begin{aligned} r\ge\min_{2\le\alpha\le r}\{\sqrt[\alpha-1]{\frac{\Delta\mu m(m\alpha-\alpha+1)+m\alpha^2 }{\alpha(\Delta\mu+1)(m\alpha-\alpha+1)}} \\ \cdot(N-\alpha+1)+\alpha-1\} \end{aligned}$\\[2mm] 
				& {\bf Probabilistic}   & $p\le \max_{\alpha\ge2}\{1-\sqrt[\alpha-1]{\frac{m}{m\alpha-\alpha+1}}\}$&    $p\le\max_{\alpha\ge2}\{1-\sqrt[\alpha-1]{\frac{m(\Delta\mu(m\alpha-\alpha+1)+\alpha^2)}{\alpha(\Delta\mu +1)(m\alpha-\alpha+1)}}\}$\\[2mm]
				\bottomrule
			\end{tabular}
	\end{center}
			\end{small}
\end{table*}

The probability of successful recovery, denoted by $P_s(\alpha)$, is another important performance metric  \cite{Sardari_Allocation_2010,leong2012distributed}. Since $P_s(\alpha)$ and $\mu_s(\alpha)$ may exhibit different trends when varying $\alpha$, we also make comparisons between these two metrics in our numerical analysis to find an overall optimal allocation. 

We here put together the relevant results of \cite{noori2016storage}, which focuses on small files, and \cite{peng2018distributed}, which focuses on large files. Moreover, we compare the results of these papers and those published in \cite{Sardari_Allocation_2010}, which focuses on the probability of access rather than the service rate as a performance metric. Further contributions include the following:
\begin{itemize}[leftmargin=*]
\item We provide a new numerical analysis for each access model for small files to characterize how DSS service rate and recovery probability change with $\alpha$ under different $r$ and $p$.
\item We consider several examples for each access model for large files. They are helpful in better understanding the allocation problem. We provide an analysis that results in finding the optimal $\alpha$ in the considered cases. 
\item We obtain the minimal spreading allocation (non) optimality  conditions for large files under each access model. 
\item We present a new numerical analysis for large files under each access model to show the tradeoffs between DSS service rate and recovery probability.
\end{itemize}
\section{Storage Allocation for small files}
\label{sec:small}
For the small file model, we assume the service time at each node follows an exponential distribution with the mean $\frac{1}{\mu}$. The service rate for an accessed set of nodes $\mathcal{A}$ is $\mu_s(\alpha|\mathcal{A})$, where $\mu_s(\alpha|\mathcal{A})>0$ if $\varphi(\mathcal{A})\ge\alpha$, and $0$ otherwise. Thus, the DSS service rate $\mu_{s}(\alpha)$ depends on all possible sets $\mathcal{A}$, which satisfy $\sum_{i \in \mathcal{A}} s_i \ge k$. 
In the following two subsections, we determine the $\mu_{s}(\alpha)$ for the two considered access models. Some of results in this section were published in \cite{noori2016storage}.

\subsection{Fixed-Size Access Model}
For the fixed-size access model and an exponential service time, the DSS service rate in \eqref{eq:service} becomes
\begin{equation}
      \mu_s(\alpha)=\frac{\mu}{\binom N r} \sum^{\min(r,\, m\alpha)}_{\varphi=\alpha}\frac{1}{H_\varphi-H_{\varphi-\alpha}} \binom {m\alpha}{\varphi} \binom {N-m\alpha}{r-\varphi}.
        \label{eq:service_exp}
\end{equation}
Using \eqref{eq:prob}, the probability of successful file recovery is
\begin{equation}
P_s(\alpha)=
 \sum^{\min(r,\, m\alpha)}_{\varphi=\alpha}
\frac{ \binom {m\alpha}{\varphi} \binom {N-m\alpha}{r-\varphi}}{\binom N r}.
\label{eq:prob_fix}
\end{equation}
Here and in the rest of the paper, use $\varphi$ instead of $\varphi(\mathcal{A})$. 
The following lemma gives the minimal spreading service rate .
\begin{lemma}\label{LE:mini_exp}
Under the fixed-size access model with an exponential service time, the service rate of minimal spreading allocation, i.e. $\alpha = 1$, is 
$\mu_{\mathrm{s}}(1) = \mu mr/{N}$.
\end{lemma}

\begin{IEEEproof}
From \eqref{eq:service_exp}, we get
\[
\mu_s(1)=\frac{\mu}{\binom N r} \sum^{\min(r,m)}_{\varphi=1}\frac{1}{H_\varphi-H_{\varphi-1}} \binom {m}{\varphi} \binom {N-m}{r-\varphi}.
\]
Notice that when $\varphi>m\alpha$, $\binom {m\alpha}{\varphi}=0$. Thus,
\begin{align*}
\mu_s(1)=\frac{\mu}{\binom N r} \sum^{r}_{\varphi=1}\varphi \binom {m}{\varphi} \binom {N-m}{r-\varphi}=\frac{\mu m}{\binom N r} \sum^{r}_{\varphi=1} \binom {m-1}{\varphi-1} \binom {N-m}{r-\varphi}.
\end{align*}
Using Vandermonde's convolution, one can show that
$\sum^{r}_{\varphi=1} \binom {m-1}{\varphi-1} \binom {N-m}{r-\varphi}= \binom {N-1} {r-1}$.
Therefore, $\mu_s(1)=\frac{\mu mr}{N}$.
\end{IEEEproof}

We next find an upper bound on $\mu_s(\alpha)$ for any $2 \leq \alpha \leq r$ and compare this bound with $\mu_{s}(1)$.

\begin{lemma}
\label{LE:bound_exp}
Under the fixed-size access model and an exponential service time, $\mu_{s}(\alpha)<\frac{\mu mr}{N}$ for $2\le\alpha\le r$.
\end{lemma}

\begin{IEEEproof}
By applying \eqref{ieq:rate_exp} to \eqref{eq:service_exp} and using Vandermonde's convolution, we arrive at
\begin{align*}
    \mu_{s}(\alpha)&<\frac{\mu}{\alpha\binom N r} \sum^{\min(r,m\alpha)}_{\varphi=\alpha}\varphi \binom {m\alpha}{\varphi} \binom {N-m\alpha}{r-\varphi}\\
    &<\frac{\mu m}{\binom N r}\sum^{r-1}_{\varphi=0} \binom {m\alpha-1}{\varphi} \binom {N-m\alpha}{r-1-\varphi}
    =\frac{\mu m\binom {N-1} {r-1}}{\binom N r}=\frac{\mu mr}{N}.
\end{align*}
\end{IEEEproof}
Lemmas~\ref{LE:mini_exp} and \ref{LE:bound_exp}, give the following theorem on the optimality of minimum spreading ($\alpha=1$).

\begin{theorem}
 \label{TH:us1}
 Under the fixed-size access and exponential service, the minimal spreading maximizes the service rate.
\end{theorem}

\begin{figure}[hbt]
	\centering
	\includegraphics[width=0.725\textwidth]{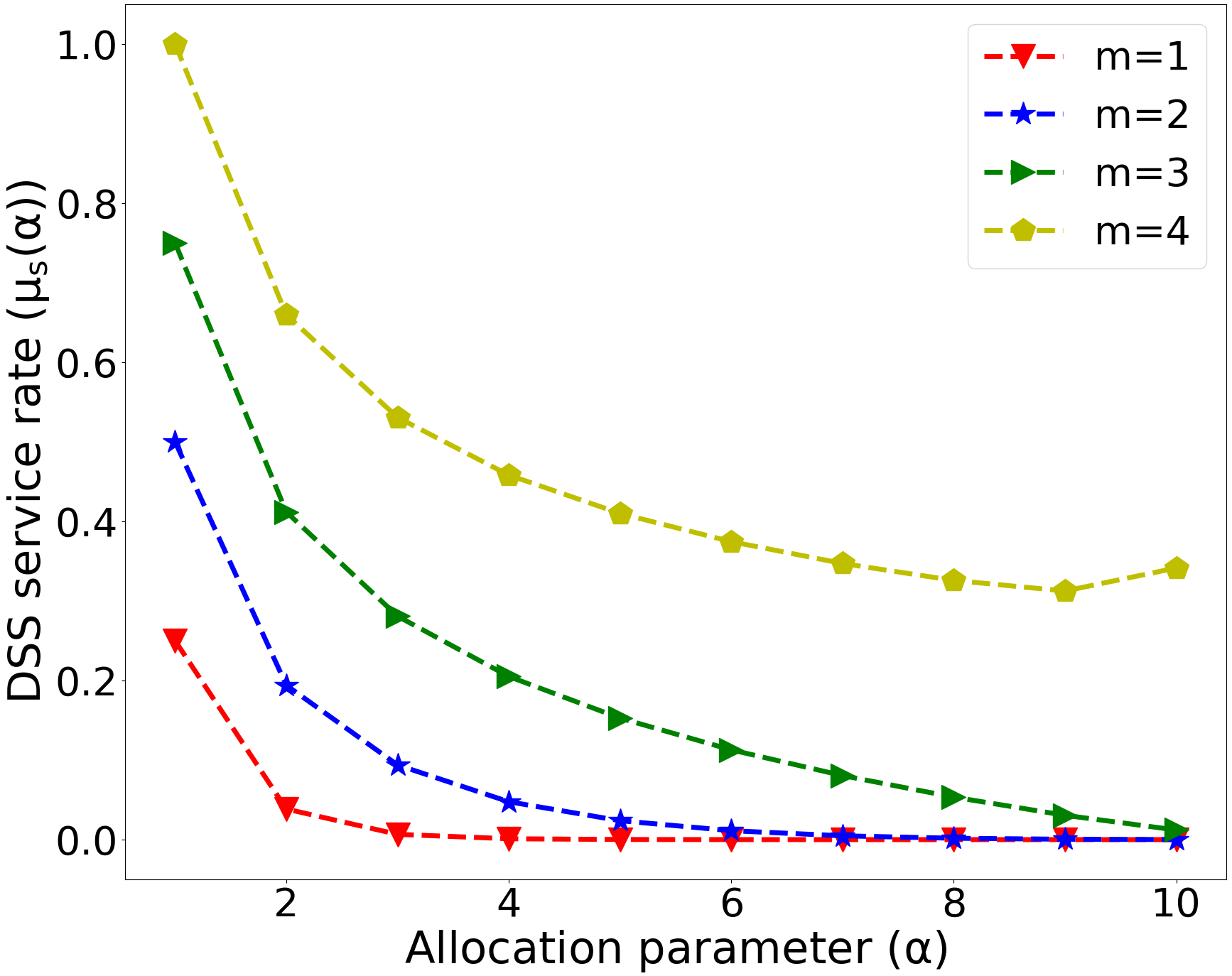}~\\
	\includegraphics[width=0.725\textwidth]{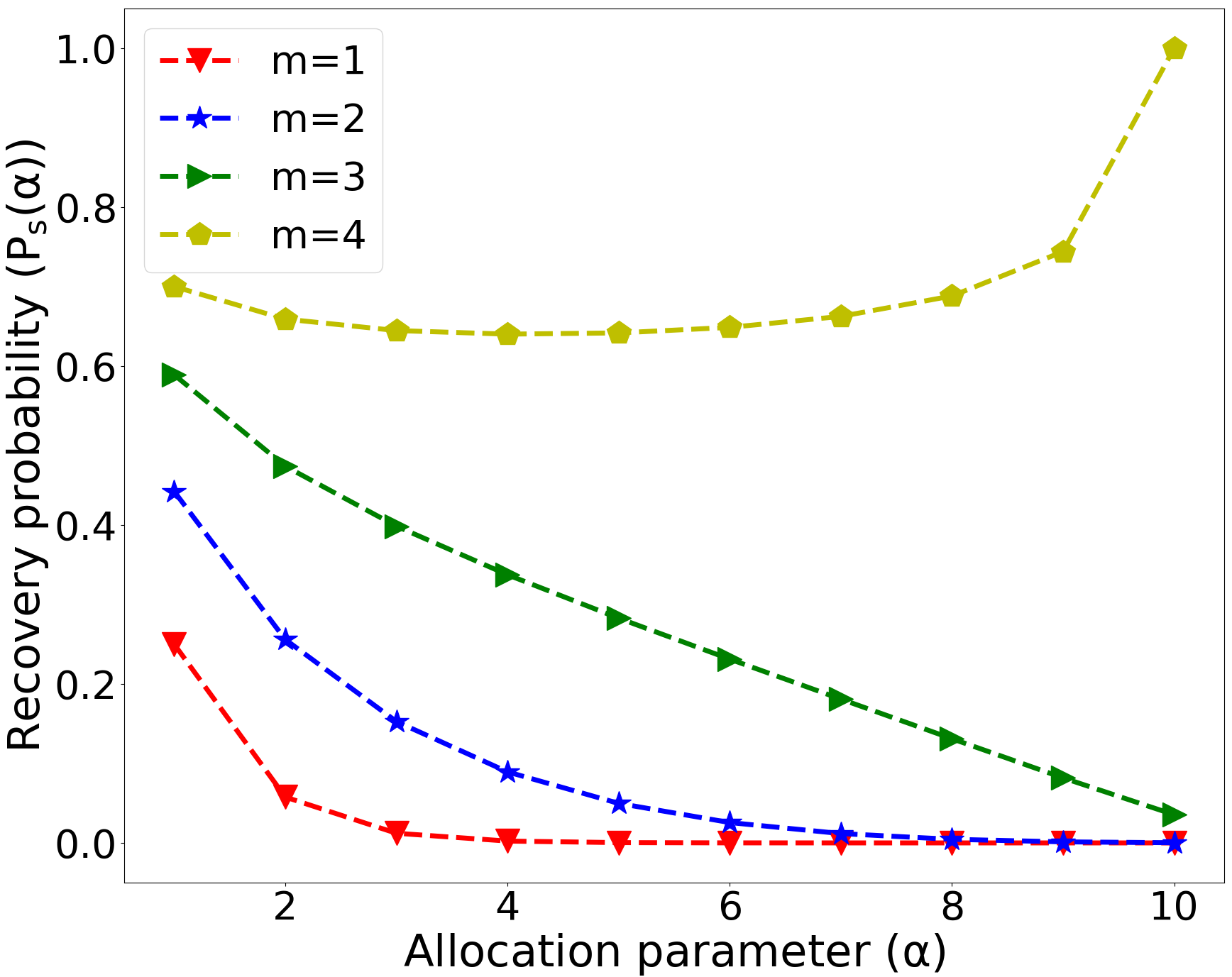}~
	\caption{ Comparing the service rate $\mu_{s}(\alpha)$ (upper, cf.\eqref{eq:service_exp}) and the successful recovery probability $P_s(\alpha)$ (lower, cf.\eqref{eq:prob_fix}) for a range of allocation parameters $\alpha$, under the fixed-size access model. The number of nodes $N$ is $40$, and the number of accessed nodes $r$ is $10$. The service time follows $\expD(1)$. When $m\le 3$, the minimal spreading allocation is optimal for both metrics. When $m=4$, the minimal spreading maximizes the service rate whereas
	the maximal spreading maximizes the probability of success recovery. }
	\label{fig:small_file_m}
\end{figure}

\paragraph*{Numerical Analysis}
In Fig.~\ref{fig:small_file_m}, we evaluate \eqref{eq:service_exp} and \eqref{eq:prob_fix} to see how the DSS service rate $\mu_{s}(\alpha)$ (upper) and the successful recovery probability $P_s(\alpha)$ (lower) changes with the allocation parameter $\alpha$. We consider a system with $N=40$ storage nodes and $r=10$ accessed nodes for four different levels of redundancy $m\in\{1,2,3,4\}$. In the upper subfigure, $\mu_{s}(\alpha)$ reaches its maximum at $\alpha=1$, i.e. the minimal spreading allocation  is optimal. When $m\le 3$, $\mu_{s}(\alpha)$ decreases with increasing $\alpha$ and approaches $0$. When $m=4$, $\mu_{s}(\alpha)$ reaches its minimum at $\alpha=9$. In the lower subfigure, when $m\le 3$, $P_s(\alpha)$ reaches its maximum at $\alpha=1$, thus the minimal spreading allocation is optimal. When $m=4$, $P_s(\alpha)$ reaches $1$ at $\alpha=10$, i.e. the maximal spreading allocation ($\alpha=N/m$) is optimal. From the observations, we conclude that the minimal spreading allocation is always optimal under the DSS service rate, which is consistent with the result in Theorem~\ref{TH:us1}. The optimal allocation under successful recovery probability is determined by the level of introduced redundancy. 

\begin{figure}[hbt]
	\centering
	\includegraphics[width=0.725\textwidth]{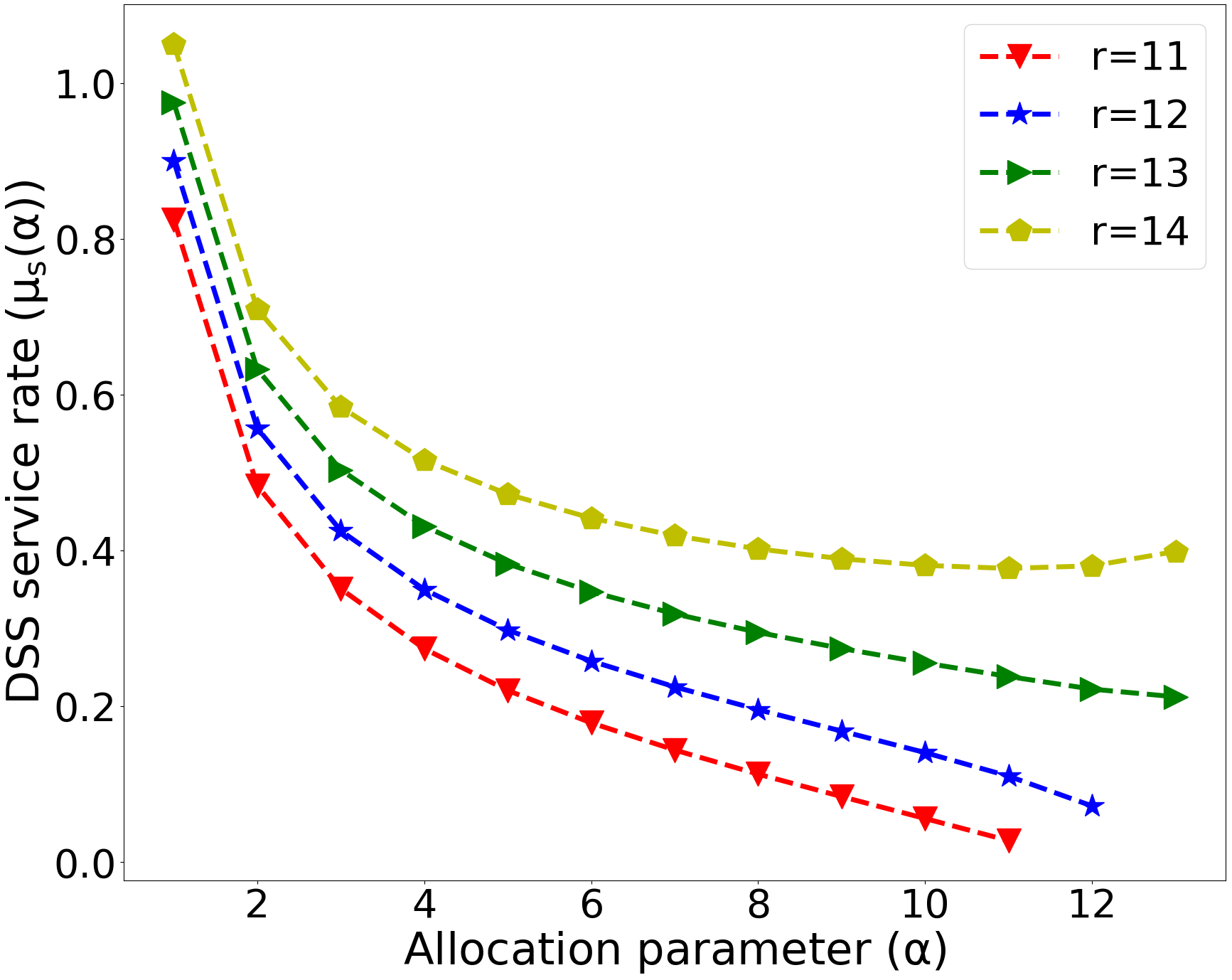}~\\
	\includegraphics[width=0.725\textwidth]{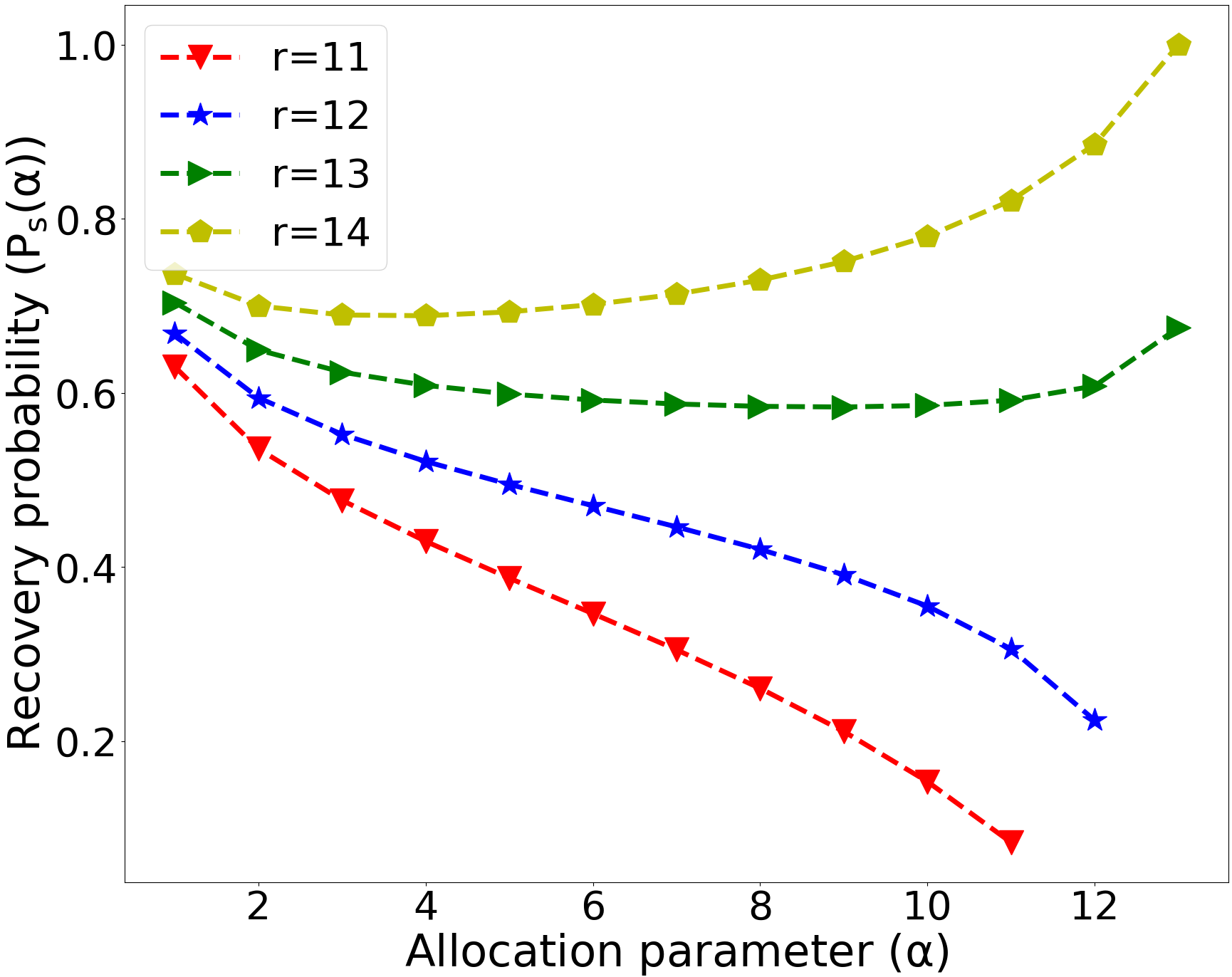}~
	\caption{ Comparing the service rate $\mu_{s}(\alpha)$ (upper, cf.~\eqref{eq:service_exp}) and the successful recovery probability $P_s(\alpha)$ (lower, cf.~\eqref{eq:prob_fix}) for a range of allocation parameters $\alpha$, under the fixed-size access model. The number of nodes $N$ is $40$, and the redundancy level $m$ is $3$. The service time follows $\expD(1)$. When $r\le 13$, the minimal spreading allocation is optimal in both figures. When $r=14$, $\alpha=12$ allocation maximizes the successful recovery probability.  }
	\label{fig:small_file_r}
\end{figure}

In Fig.~\ref{fig:small_file_r}, we analyze $\mu_{s}(\alpha)$ vs. $\alpha$ (upper) and $P_s(\alpha)$ (lower) vs. $\alpha$ for different numbers of accessed node $r\in\{10,11,12,13\}$. We consider a system with $N=40$ storage nodes and a redundancy level of $m=3$. In the upper subfigure, $\mu_{s}(\alpha)$ reaches its maximum at $\alpha=1$, i.e. the minimal spreading allocation is optimal. $\mu_{s}(\alpha)$ decreases with increasing $\alpha$ except for the scenario $r=14$. In the lower subfigrue,  when $r\le 13$, $P_s(\alpha)$ reaches its maximum at $\alpha=1$, i.e. the minimal spreading allocation is optimal. When $r=14$, $P_s(\alpha)$ reaches its maximum at $\alpha=12$. From the observations, we conclude that the minimal spreading allocation is always optimal under the DSS service rate. However, it is no longer optimal under the successful recovery probability as $r$ increases. 

\subsection{Probabilistic Access Model}
For probabilistic access model under exponential service time, the DSS service rate \eqref{eq:service} becomes
\begin{equation}
      \mu_s(\alpha)=\sum^{m\alpha}_{\varphi=\alpha}\frac{\mu}{H_\varphi-H_{\varphi-\alpha}} \binom {m\alpha}{\varphi} (1-p)^{\varphi}p^{m\alpha-\varphi}.
        \label{eq:service_prob_exp}
\end{equation}
Using \eqref{eq:prob}, the probability of successful file recovery is
\begin{equation}
P_s(\alpha)=
 \sum^{\min(r,\, m\alpha)}_{\varphi=\alpha}
\binom {m\alpha}{\varphi} (1-p)^{\varphi}p^{m\alpha-\varphi}.
\label{eq:prob_prob}
\end{equation}
We have the following result on the minimal spreading.
\begin{lemma}\label{LE:mini_prob_exp}
Under the probabilistic access model with exponential service time, the service rate of the minimal spreading allocation,  i.e. $\alpha = 1$, is $\mu_{\mathrm{s}}(1) = \mu m (1-p)$.
\end{lemma}

\begin{IEEEproof}
From \eqref{eq:service_prob_exp}, we get
\begin{align*}
    \mu_{s}(1)
    =\mu m(1-p)\sum^{ m-1}_{\varphi=0} \binom { m-1}{\varphi} (1-p)^{\varphi}p^{ m-\varphi-1}
\end{align*}
Using binomial expansion, we get $\mu_{s}(1)= \mu m (1-p)$.
\end{IEEEproof}
Similar to Lemma~\ref{LE:bound_exp}, we find an upper bound on the DSS service rate when $\alpha\ge2$.

\begin{lemma}
\label{LE:bound_prob_exp}
Under the probabilistic access model with an exponential service time, for any $\alpha$ quasi-uniform allocation, its service rate satisfies $\mu_{s}(\alpha)<\mu m (1-p)$.
\end{lemma}

\begin{IEEEproof}
By applying \eqref{ieq:rate_exp} to \eqref{eq:service_prob_exp}, we arrive at
\begin{align*}
    &\mu_{s}(\alpha)<
    \mu m (1-p)\sum^{m\alpha-1}_{\varphi=\alpha-1} \binom {m\alpha-1}{\varphi} (1-p)^{\varphi}p^{m\alpha-\varphi-1}<\\
    &\mu m (1-p)\sum^{m\alpha-1}_{\varphi=0} \binom {m\alpha-1}{\varphi} (1-p)^{\varphi}p^{m\alpha-\varphi-1}
    =\mu m (1-p).
\end{align*}
\end{IEEEproof}
Using Lemmas~\ref{LE:mini_prob_exp} and \ref{LE:bound_prob_exp}, we have the following result on the optimal storage allocation for the probabilistic access model.

\begin{theorem}
 \label{TH:us1_prob}
 Under the probabilistic access model with an exponential service time, the minimal spreading allocation  maximizes the DSS service rate.
\end{theorem}

\begin{figure}[hbt]
	\centering
	\includegraphics[width=0.725\textwidth]{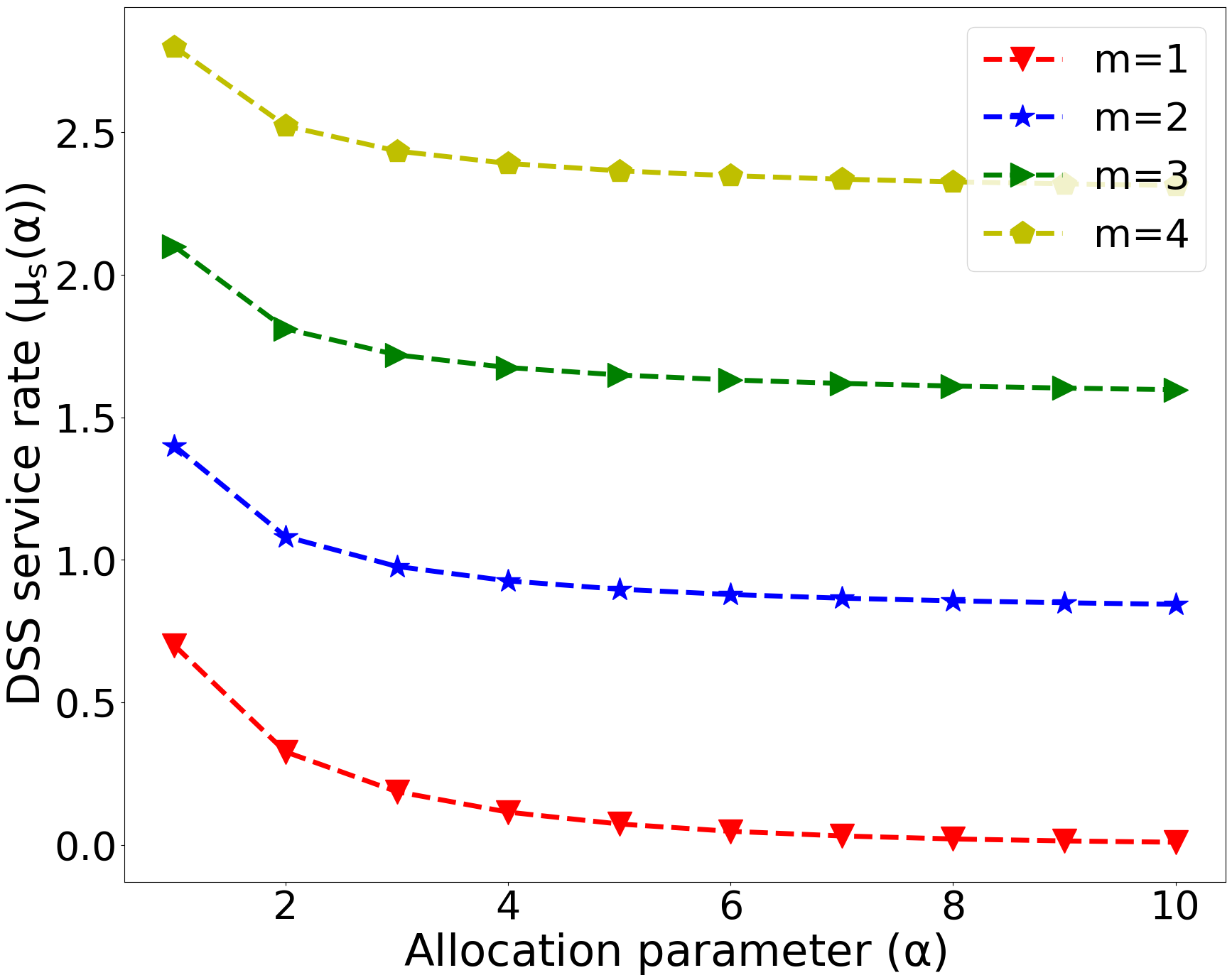}~\\
	\includegraphics[width=0.725\textwidth]{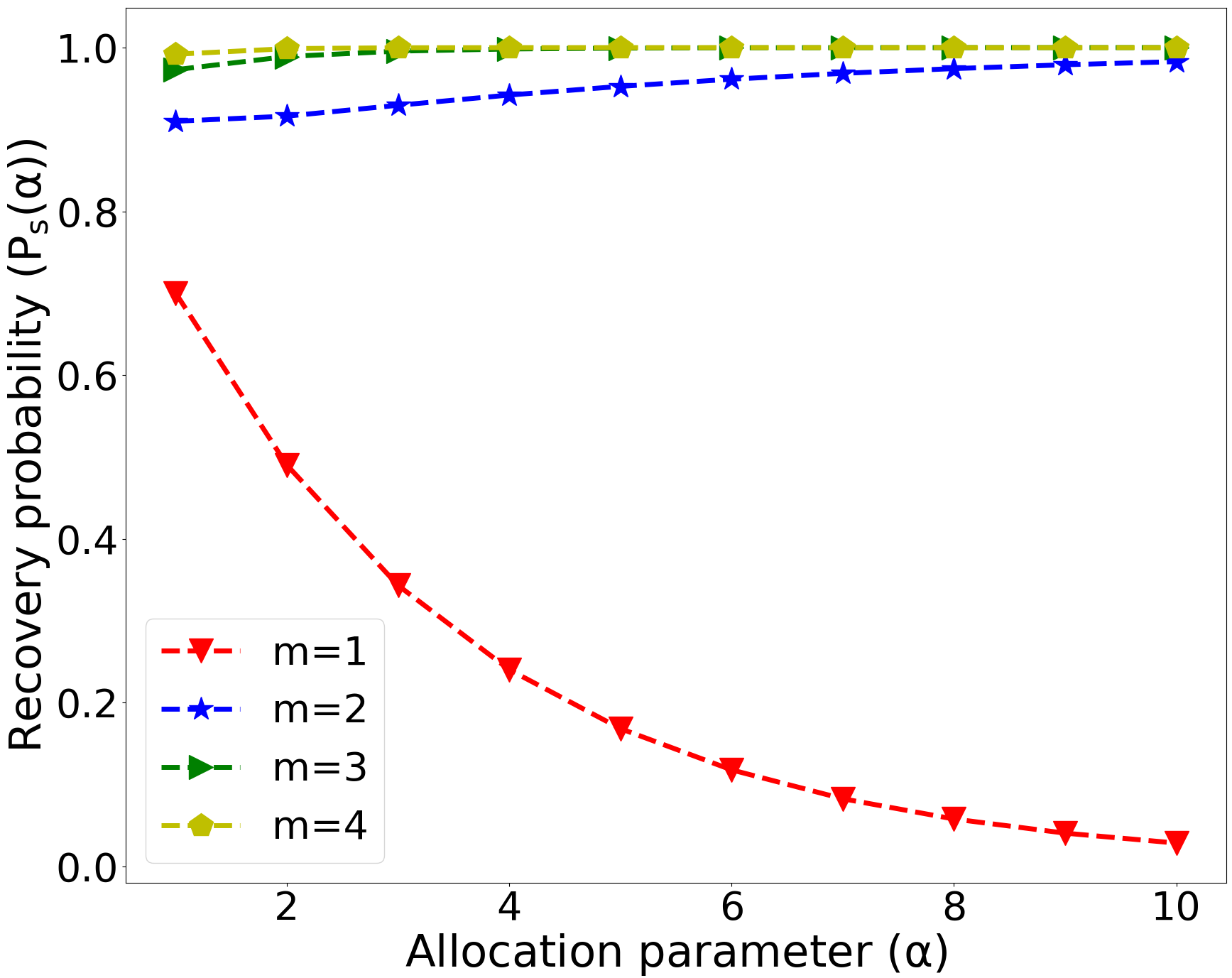}~
	\caption{ Comparisons between the service rate $\mu_{s}(\alpha)$ (upper, cf.~\eqref{eq:service_prob_exp}) and the successful recovery probability $P_s(\alpha)$ (lower, cf.~\eqref{eq:prob_prob}) as a function of the allocation parameter $\alpha$ under the probabilistic access model. The number of storage nodes is $N=40$, and the probability of failed access is $p=0.3$. The service time follows $\expD(1)$. When $m= 1$, the minimal spreading allocation is optimal in both figure. When $m\ge2$, the minimal spreading allocation is optimal considering $\mu_{s}(\alpha)$, and performs the worst considering $P_s(\alpha)$.  }
	\label{fig:small_file_pm}
\end{figure}

\paragraph*{Numerical Analysis}
In Fig.~\ref{fig:small_file_pm}, we respectively evaluate \eqref{eq:service_prob_exp} and  \eqref{eq:prob_prob} to see how the DSS service rate $\mu_{s}(\alpha)$ (upper) and the successful recovery probability $P_s(\alpha)$ (lower) changes with the allocation parameter $\alpha$. We consider a system with $N=40$ storage nodes and a failed access probability $p=0.3$ for four different levels of redundancy $m\in\{1,2,3,4\}$. In the upper subfigure, $\mu_{s}(\alpha)$ reaches its maximum at $\alpha=1$, i.e. the minimal spreading allocation is optimal, and decreases with increasing $\alpha$. In the lower subfigure, when $m=1$, $P_s(\alpha)$ reaches its maximum at $\alpha=1$, i.e. the minimal spreading allocation is optimal. When $m\ge2$, $P_s(\alpha)$ reaches its maximum at $\alpha=10$, approaching $1$ for $m=3$ and $4$ when $\alpha\ge4$. From the observations, we conclude that the minimal spreading allocation is always optimal under the DSS service rate, which is consistent with the result in Theorem~\ref{TH:us1_prob}. The optimal allocation under successful recovery probability is determined by the level of introduced redundancy. 

\begin{figure}[hbt]
	\centering
	\includegraphics[width=0.725\textwidth]{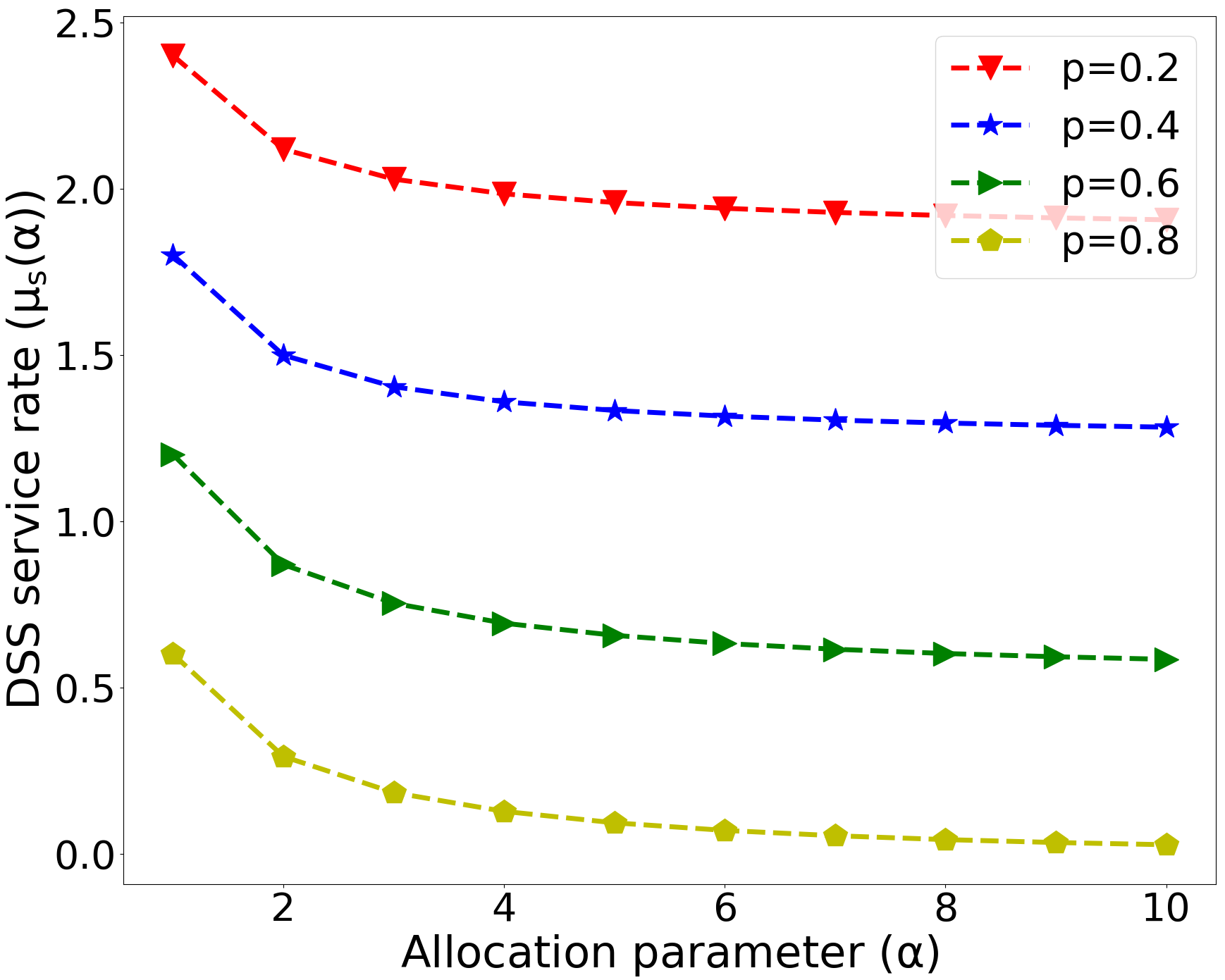}~\\
	\includegraphics[width=0.725\textwidth]{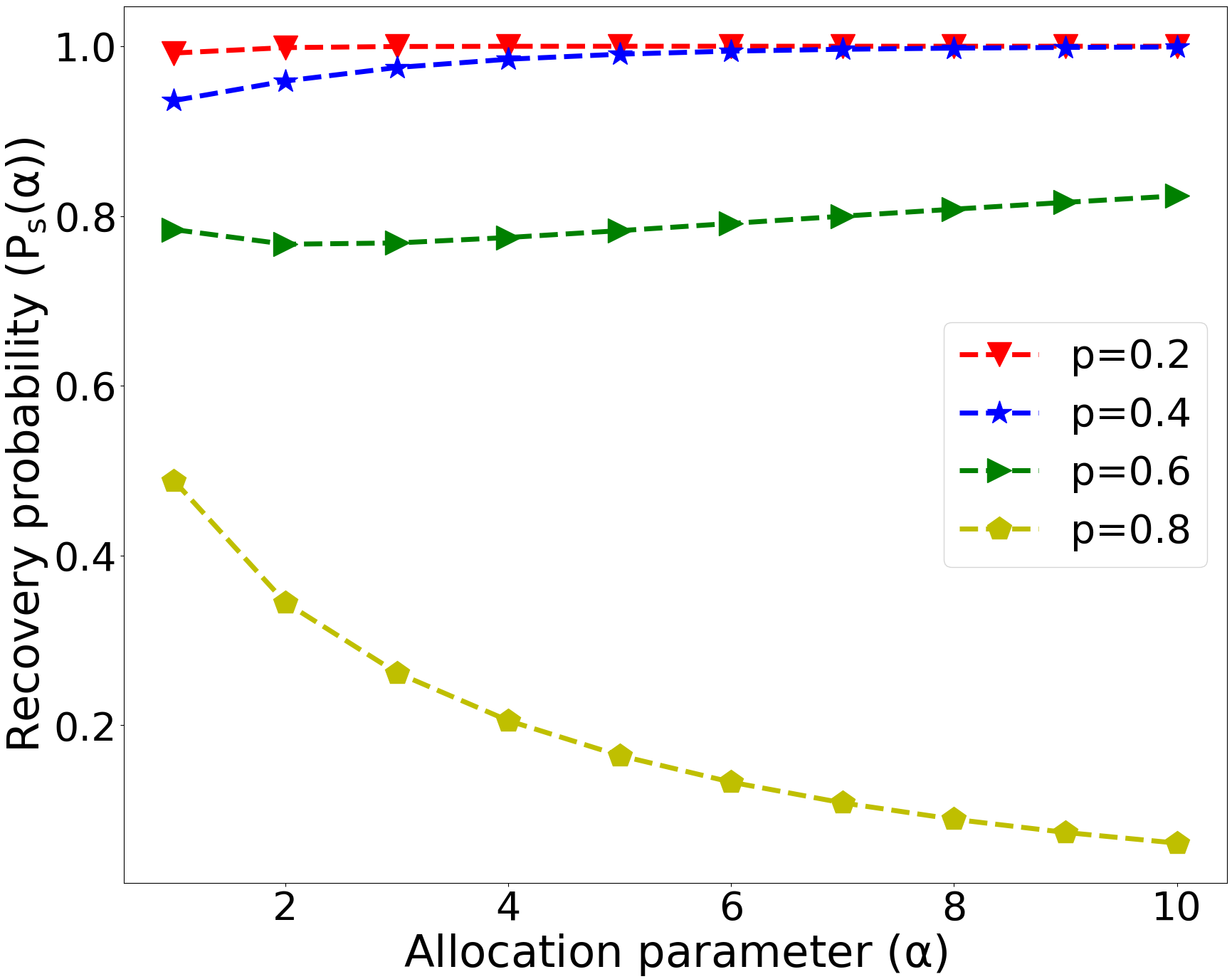}~
	\caption{ Comparisons between the DSS service rate $\mu_{s}(\alpha)$ (upper, cf.~\eqref{eq:service_prob_exp}) and the successful recovery probability $P_s(\alpha)$ (lower, cf.~\eqref{eq:prob_prob}) as a function of the allocation parameter $\alpha$ under the probabilistic access model. The number of storage nodes is $N=40$, and the redundancy level is $m=3$. The service time follows $\expD(1)$. minimal spreading allocation  is optimal with regards to $\mu_{s}(\alpha)$. For $p=0.8$, the minimal spreading allocation  is also optimal for $P_s(\alpha)$ while it performs the worst when $p\le 0.4$.}
	\label{fig:small_file_pp}
\end{figure}

In Fig.~\ref{fig:small_file_pp}, we analyze $\mu_{s}(\alpha)$ vs. $\alpha$ (upper) and $P_s(\alpha)$ (lower) vs. $\alpha$  for $p\in\{0.2,0.4,0.6,0.8\}$. We consider a system with $N=40$ storage nodes and a redundancy level of  $m=3$. In the upper subfigure, $\mu_{s}(\alpha)$ reaches its maximum at $\alpha=1$, i.e. the minimal spreading allocation  is optimal. $\mu_{s}(\alpha)$ decreases with increasing $\alpha$. In the lower subfigrue, when $p\le 0.4$, $P_s(\alpha)$ increases with $\alpha$, and reaches $1$ at about $\alpha\ge6$. When $p=0.6$, $P_s(\alpha)$ takes values around $0.8$. When $p=0.8$, $P_s(\alpha)$ reaches its maximum at $\alpha=1$. From the observations in Fig.~\ref{fig:small_file_pp}, we conclude that the minimal spreading allocation  is always optimal under the DSS service rate. However, the optimal allocation under the successful recovery probability depends on the probability of failed access.

\section{Storage Allocation for Large Files with Scaled Exponential Service Time}
\label{sec:large_exp}
For the large file model, we assume that the service time at each node follows a scaled exponential distribution with the mean $\frac{1}{\alpha\mu}$, i.e. the allocation parameter $\alpha$ changes the scale of an exponential distribution. In the following two subsections, we determine the $\mu_{s}(\alpha)$ for the two considered access models. Some of the results in this section were published in \cite{peng2018distributed}.

\subsection{Fixed-Size Access Model}
For the fixed-size access model under a scaled exponential service time, the DSS service rate \eqref{eq:service} becomes
\begin{equation}
      \mu_s(\alpha)=\frac{\mu\alpha}{\binom N r} \sum^{\min(r,m\alpha)}_{\varphi=\alpha}\frac{1}{H_\varphi-H_{\varphi-\alpha}} \binom {m\alpha}{\varphi} \binom {N-m\alpha}{r-\varphi}.
        \label{eq:service_scale}
\end{equation}
By comparing \eqref{eq:service_exp} and \eqref{eq:service_scale}, we expect the minimal spreading to not always be optimal. We prove that the maximal spreading performs better under some system parameters' values.

\begin{lemma}
\label{Le:spreading_fix}
Under fixed-size access model with a scaled exponential service time, the service rate of the maximal spreading allocation, i.e. $\alpha=r$, is $\mu_s(r)= \frac{\mu r \binom{rm}{r}}{H_r\binom{N}{r}}$.
The service rate of the minimal spreading allocation, i.e. $\alpha=1$, is $\mu_s(1)= \frac{\mu m\binom{N-1}{r-1}}{\binom{N}{r}}=\frac{\mu m r}{N}$.
\end{lemma}
\begin{IEEEproof}
From \eqref{eq:service_scale}, for the maximal spreading allocation, we get {$
\mu_s(r)=\frac{\mu r}{\binom N r} \sum^{\min(r,r m)}_{\varphi=r}\frac{1}{H_\varphi-H_{\varphi-r}} \binom {r m}{\varphi}\cdot \binom {N-r m}{r-\varphi}$}.
Since $m\ge1$ and $H_0=0$, { $\mu_s(r)=\frac{\mu r}{\binom N r}\frac{1}{H_r} \binom {r m}{r} \binom {N-r m}{r-r}=\frac{\mu r \binom{rm}{r}}{H_r\binom{N}{r}}$}.

For the minimal spreading allocation, we get
{ $\mu_s(1|\varphi)=\mu \varphi$}.
Thus, by applying the same approach as in the proof of Lemma~\ref{LE:mini_exp}, we get { $\mu_s(1)= \frac{\mu m r}{N}=\frac{\mu m\binom{N-1}{r-1}}{\binom{N}{r}}$}.
\end{IEEEproof}

With Lemma~\ref{Le:spreading_fix}, we obtain the following result.

\begin{theorem}[minimal vs.\ maximal spreading]
\label{Le:fixed_spreading_scale}
Under the fixed-size access model with a scaled exponential service time, when $rm\ge N$, the maximal spreading allocation outperforms the minimal spreading allocation, i.e. $\mu_s(r)\ge\mu_s(1)$.
\end{theorem}
\begin{IEEEproof}
Lemma~\ref{Le:spreading_fix} with $rm\ge N$ and $r\ge H_r$ gives\\
$\mu_s(r)=\frac{\mu r \binom{rm}{r}}{H_r\binom{N}{r}}=\frac{\mu rm \binom{rm-1}{r-1}}{H_r\binom{N}{r}} \ge\frac{\mu m H_r \binom{N-1}{r-1}}{H_r\binom{N}{r}}=\frac{\mu m r}{N}=\mu_s(1)$.
\end{IEEEproof}

\begin{remark}
Under the fixed-size access model with a scaled exponential service time, the minimal spreading allocation does not always maximize the DSS service rate.
 \label{Th:fixed_scale}
\end{remark}

\subsubsection{Optimal and Non-optimality Conditions for the Minimal Spreading Allocation}\label{subsubsec:condition1}
Considering the complexity of \eqref{eq:service_scale}, finding the $\alpha$ that maximizes $\mu_s(\alpha)$ is hard. Instead, we find the optimality and non-optimality conditions for the minimal spreading allocation in the following. 

\begin{theorem}[minimal spreading optimality condition]
 Under the fixed-size access with a scaled exponential service, the minimal spreading  maximizes the service rate $\mu_s(\alpha)$ when\\ { $r\le \min_{2\le \alpha \le r} \{1+\frac{N-1}{\sqrt[\alpha-1]{\alpha \binom{m\alpha-1}{\alpha-1}}}\}$}.
 \label{Th:fix_opt_scale}
\end{theorem}

\begin{IEEEproof}
We need to find the condition ensuring $\mu_s(1)\ge \mu_s(\alpha)$ for all $2\le \alpha\le r$. 
Consider $\mu_s(\alpha)$ in \eqref{eq:service_scale}. By  \eqref{ieq:rate}, we have $\mu_s(\alpha|\varphi)<\mu\varphi$ when $\alpha\ge2$, and thus
\begin{align*}
    \mu_s(\alpha)&<\frac{\mu}{\binom{N}{r}}\sum^{\min(r,m\alpha)}_{\varphi=\alpha}\varphi \binom{m\alpha}{\varphi}\binom{N-m\alpha}{r-\varphi}\\
      &=\frac{\mu m\alpha}{\binom{N}{r}}\sum^{\min(r,m\alpha)}_{\varphi=\alpha} \prod^{\alpha-2}_{i=0}\frac{m\alpha-1-i}{\varphi-1-i} \binom{m\alpha-\alpha}{\varphi-\alpha}\binom{N-m\alpha}{r-\varphi}.
    \end{align*}
Since $\varphi$ goes from $\alpha$ to $m\alpha$, we further have 
    \begin{align*}
    \mu_s(\alpha)
    &<\frac{\mu m\alpha}{\binom{N}{r}}\sum^{\min(r,m\alpha)}_{\varphi=\alpha}(\prod^{\alpha-2}_{i=0}\frac{m\alpha-1-i}{\alpha-1-i}) \binom{m\alpha-\alpha}{\varphi-\alpha}\binom{N-m\alpha}{r-\varphi}\\
   & =\frac{\mu m\alpha\binom{m\alpha-1}{\alpha-1} \binom{N-\alpha}{r-\alpha}}{\binom{N}{r}} ~
    \text{(by Vandermonde's\ convolution)}.
    \end{align*}
According to Lemma~\ref{Le:fixed_spreading_scale}, we have
$\mu_s(1)=\frac{\mu m\binom{N-1}{r-1}}{\binom{N}{r}}$.
To satisfy $\mu_s(\alpha)\le \mu_s(1)$, we have
\begin{equation}
\begin{split}
   \frac{\mu m\alpha\binom{m\alpha-1}{\alpha-1} \binom{N-\alpha}{r-\alpha}}{\binom{N}{r}}\le\frac{\mu m\binom{N-1}{r-1}}{\binom{N}{r}}
   \Leftrightarrow \alpha \binom{m\alpha-1}{\alpha-1}\le\prod^{\alpha-1}_{i=1}\frac{N-i}{r-i}.
    \end{split}
\label{eq:scaled_fix}
\end{equation}
As $\frac{N-i}{r-i}<\frac{N-1-i}{r-1-i}$ for $N>r$, it can be shown that
$\prod^{\alpha-1}_{i=1}\frac{N-i}{r-i}>(\frac{N-1}{r-1})^{\alpha-1}$, and as a result, \eqref{eq:scaled_fix} holds when
\begin{equation}
\begin{split}
    &\alpha \binom{m\alpha-1}{\alpha-1}\le(\frac{N-1}{r-1})^{\alpha-1}
    \Leftrightarrow r\le 1+\frac{N-1}{\sqrt[\alpha-1]{\alpha \binom{m\alpha-1}{\alpha-1}}}.
\end{split}
\label{eq:scaled-fix2}
\end{equation}
\noindent If \eqref{eq:scaled-fix2} holds for all $2\le \alpha \le r$, $\mu_s(1)$ is optimal.
\end{IEEEproof}



\begin{theorem}[minimal spreading non-optimality condition]
 Under the fixed-size access model with a scaled exponential service time, the minimal spreading allocation does not maximize the DSS service rate $\mu_s(\alpha)$ when \\{ $ r\ge\min_{2\le\alpha\le r}\{\sqrt[\alpha-1]{\frac{m}{m\alpha-\alpha +1}}(N-\alpha+1)+\alpha-1\}$}.
 \label{Th:fix_nonopt_scale}
\end{theorem}

\begin{IEEEproof}
To prove the theorem statement, we need to find the condition ensuring $\mu_s(1)\le \mu_s(\alpha)$ for at least one $\alpha \in [2,r]$. The expression of $\mu_s(\alpha)$ is given in \eqref{eq:service_scale}.
According to \eqref{ieq:rate}, we have { $\mu_s(\alpha|\varphi)>\mu(\varphi-\alpha+1)$} when $\alpha\ge2$. Thus,
\begin{align*}
    &\mu_s(\alpha)>\frac{\mu}{\binom{N}{r}}\sum^{\min(r,m\alpha)}_{\varphi=\alpha}(\varphi-\alpha+1) \binom{m\alpha}{\varphi}\binom{N-m\alpha}{r-\varphi}\\
    &=\sum^{\min(r,m\alpha)}_{\varphi=\alpha}\frac{\mu(\varphi-\alpha+1)}{\binom{N}{r}} \prod^{\alpha-2}_{i=0}\frac{m\alpha-i}{\varphi-i}\binom{m\alpha-\alpha+1}{\varphi-\alpha+1}\binom{N-m\alpha}{r-\varphi}.
    \end{align*}
 Since\ $\varphi$ goes from $\alpha$ to $m\alpha$, we further have 
    \begin{align*}
\mu_s(\alpha)
    &>\frac{\mu}{\binom{N}{r}}\sum^{\min(r,m\alpha)}_{\varphi=\alpha}(\varphi-\alpha+1)\binom{m\alpha-\alpha+1}{\varphi-\alpha+1}\binom{N-m\alpha}{r-\varphi}\\
    &=\frac{\mu(m\alpha-\alpha+1)\binom{N-\alpha}{r-\alpha}}{\binom{N}{r}}~
    \text{(Vandermonde's convolution)}.
\end{align*}
According to Lemma~\ref{Le:fixed_spreading_scale}, we have
$   \mu_s(1)=\frac{\mu m\binom{N-1}{r-1}}{\binom{N}{r}}
$, hence, to satisfy $\mu_s(\alpha)\ge\mu_s(1)$, we need to have
\begin{equation}
\begin{split}
   &\frac{\mu(m\alpha-\alpha+1)\binom{N-\alpha}{r-\alpha}}{\binom{N}{r}}\ge\frac{\mu m\binom{N-1}{r-1}}{\binom{N}{r}}\\
   &\Leftrightarrow \frac{m\alpha-\alpha +1}{m}\ge \prod^{\alpha-2}_{i=0}\frac{N-1-i}{r-1-i}
\end{split}
\label{eq:scaled-fix3}
\end{equation}
Since { $\frac{N-1-i}{r-1-i}<\frac{N-2-i}{r-2-i}$} for $N>r$, we have
{ $\prod^{\alpha-2}_{i=0}\frac{N-1-i}{r-1-i}<(\frac{N-\alpha+1}{r-\alpha+1})^{\alpha-1}$. Hence, if $\frac{m\alpha-\alpha +1}{m}\ge(\frac{N-\alpha+1}{r-\alpha+1})^{\alpha-1}
    \Leftrightarrow  r\ge\sqrt[\alpha-1]{\frac{m}{m\alpha-\alpha +1}}(N-\alpha+1)+\alpha-1$},
the inequality \eqref{eq:scaled-fix3} holds, meaning that $\mu_s(\alpha)\ge\mu_s(1)$ and $\mu_s(1)$ is not optimal.
\end{IEEEproof}


From Theorems~\ref{Th:fix_opt_scale} and \ref{Th:fix_nonopt_scale}, we see that both conditions depend on the number of accessed nodes $r$. Roughly speaking, when $r$ is small, $\alpha=1$ is optimal while an $\alpha>1$ is optimal when $r$ is large. 
Based on these theorems, we conjecture the following about the value of $\alpha$ that maximizes $\mu_s(\alpha)$. 
\begin{conjecture}
The following assertions are true:
\begin{enumerate}
\item Under the fixed-size access model, given the number of nodes $N$ and the redundancy level $m$, the optimal $\alpha$ increases as $r$ increases. 
\item For every $N$ and $m$, there exists a $\gamma\in(1,N)$, such that for all $r\le\gamma$, the minimal spreading is optimal. 
\item For every $N$ and $m$, there exists a $\zeta\in (1,N)$, such that for all $r\ge\zeta$, the maximal spreading is optimal. 
\end{enumerate}
\label{hyp:fix1}
\end{conjecture}

\subsubsection{Numerical Analysis}
\begin{figure}[hbt]
	\centering
	\includegraphics[width=0.725\textwidth]{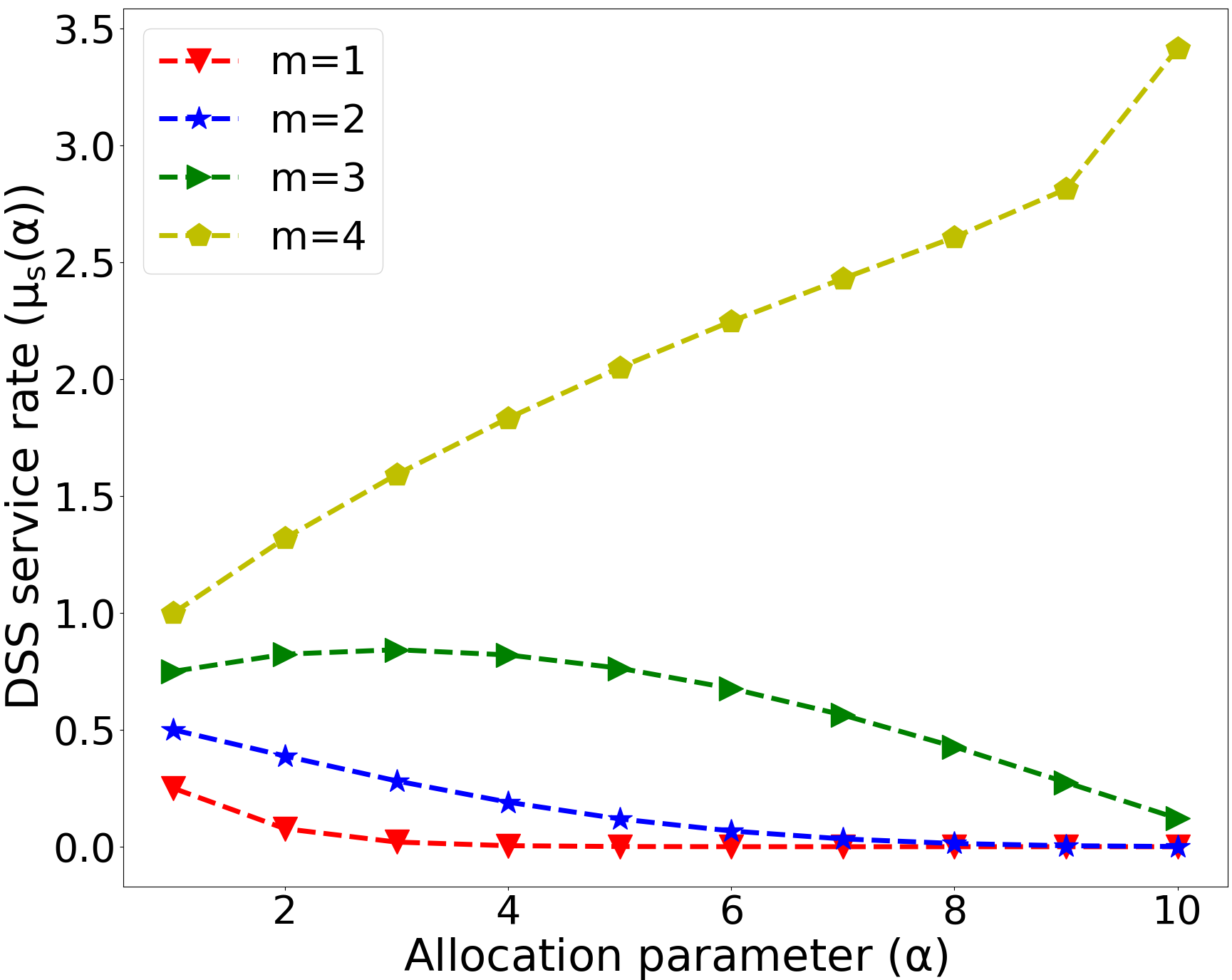}~\\
	\includegraphics[width=0.725\textwidth]{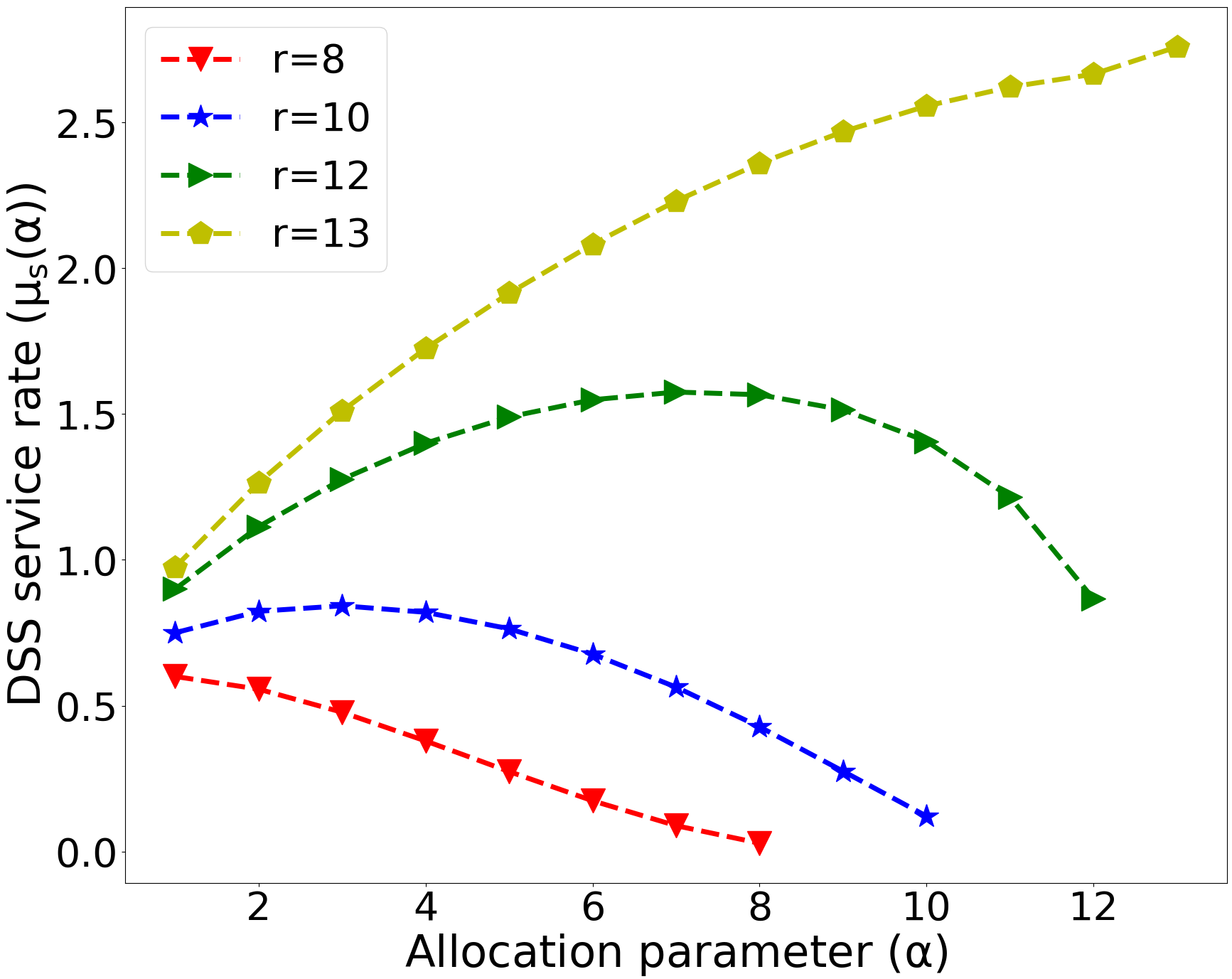}~
	\caption{ The DSS service rate $\mu_{s}(\alpha)$ for the fixed-size access model with scaled exponential service time as a function of the allocation parameter $\alpha$ (cf.~\eqref{eq:service_scale}).  The number of storage nodes is $N=40$, and the service time follows $\expD(1/\alpha)$. (upper) $\mu_{s}(\alpha)$ vs.\ $\alpha$ with $r=10$ accessed nodes for four values of $m$. (lower) $\mu_{s}(\alpha)$ vs. $\alpha$ with $m=3$ redundancy for four values of $r$. Given $r$ (or $m$), the optimal allocation changes from the minimal spreading allocation to the maximal spreading allocation as $m$ (or $r$) increases.}
	\label{fig:large_scale_fix_mr}
\end{figure}

In Fig.~\ref{fig:large_scale_fix_mr}, we evaluate the expression of $\mu_{s}(\alpha)$ given in \eqref{eq:service_scale} to see how the DSS service rate changes with the allocation parameter $\alpha$. We consider a system with $N=40$ storage nodes. 
Using Theorems~\ref{Th:fix_opt_scale} and \ref{Th:fix_nonopt_scale}, we can easily calculate the optimality and non-optimality conditions. For example, when $m=2$, the minimal spreading allocation is optimal when $r\le 7.5$ and is non-optimal when $r>27$. These conditions provide some knowledge of the optimal allocation and further insight on the optimal allocation can be found from Fig.~\ref{fig:large_scale_fix_mr}.
The upper graph shows $\mu_{s}(\alpha)$ vs. $\alpha$ for four different level of redundancy $m\in\{1, 2, 3, 4\}$, and the number of accessed nodes $r=10$. The lower graph shows $\mu_{s}(\alpha)$ vs. $\alpha$ for different numbers of accessed nodes $r\in\{8, 10, 12, 13\}$, and the redundancy level $m=3$. 
In the upper subfigure, when $m\le2$, $\mu_{s}(\alpha)$ reaches its maximum at $\alpha=1$ and decreases with increasing $\alpha$, i.e. the minimal spreading allocation is optimal. When $m=3$, $\mu_{s}(\alpha)$ reaches its maximum at $\alpha=3$. When $m=4$, $\mu_{s}(\alpha)$ increases with $\alpha$ and reaches its maximum at $\alpha=10$, i.e. the maximal spreading allocation ($\alpha=N/m$) is optimal. In the lower subfigure, when $r=8$, the minimal spreading allocation is optimal, while this is not the case for $r\ge10$. Since the redundancy level is $m=3$, which means $N/m$ is not an integer, we cannot apply the maximal spreading allocation. When $r=13$, the optimal allocation is at $\alpha=13$, where we allocate the file into $39$ of $40$ nodes. From the observations in Fig.~\ref{fig:large_scale_fix_mr}, we conclude that the minimal spreading allocation is optimal only when $m$ or $r$ is sufficiently small. By increasing either of the parameters, an $\alpha\ge2$ quasi-uniform allocation becomes optimal, and when $m$ or $r$ is sufficiently large, the maximal spreading allocation is optimal.

\begin{figure}[hbt]
\centering
\includegraphics[width=0.725\textwidth]{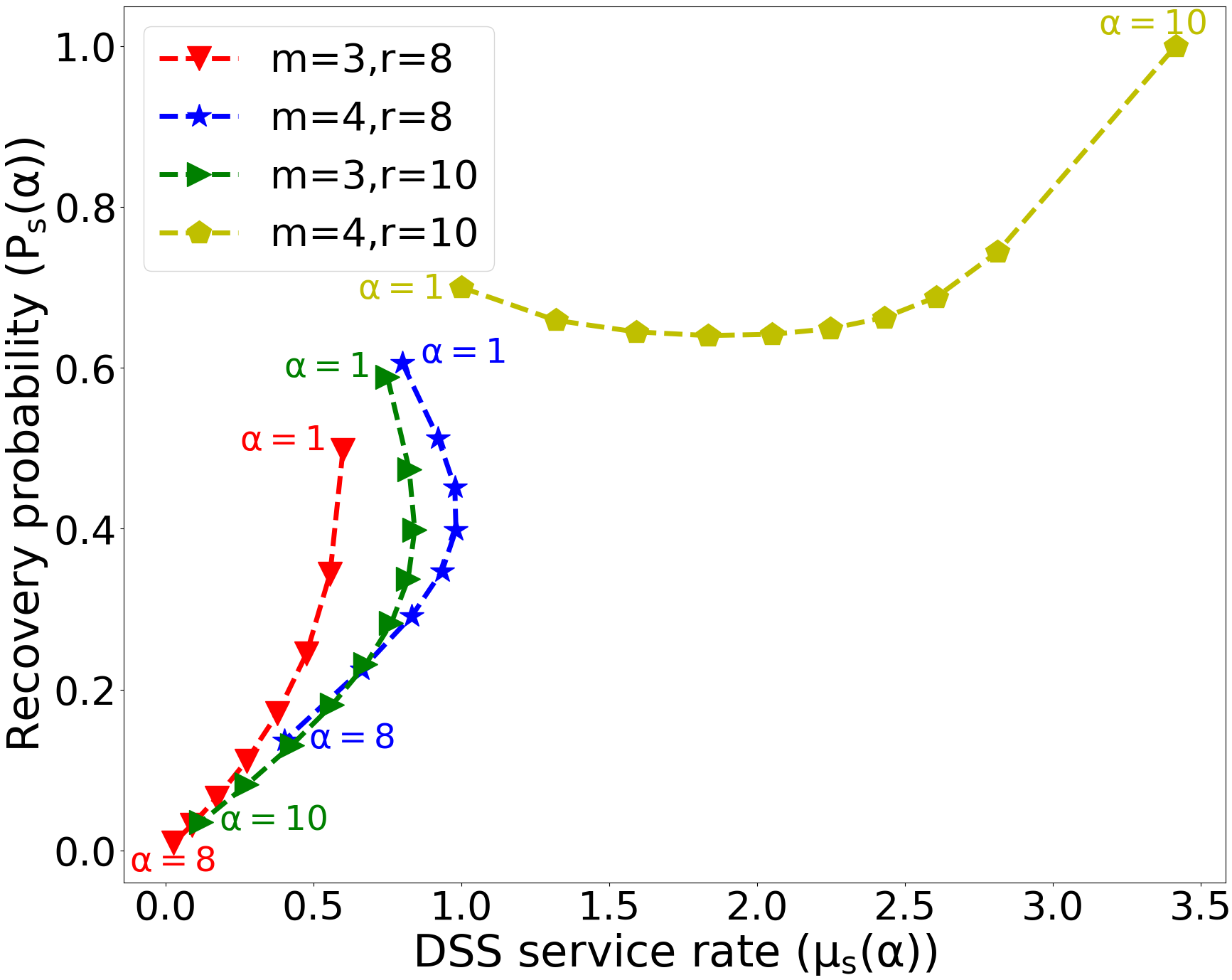}
	\caption{ Successful recovery probability $P_s(\alpha)$ vs.\ the DSS service rate $\mu_{s}(\alpha)$ for the fixed-size access model as a function of $\alpha$ for different values of $m$ and $r$ (cf.~\eqref{eq:prob_fix} and \eqref{eq:service_scale}). The number of storage nodes is $N=40$, and the service time follows $\expD(1/\alpha)$. The minimal (or maximal) spreading allocation is optimal when $m$ and $r$ are sufficiently small (or large). Otherwise, the optimal allocation is different for different performance metrics. \label{fig:large_scale_fix_up} }
\end{figure}

In Fig.~\ref{fig:large_scale_fix_up}, we analyze $P_s(\alpha)$ vs. $\mu_{s}(\alpha)$ as $\alpha$ increases from $1$ to $r$. We consider a system with $N=40$ storage nodes and two values for each parameter, i.e. $m\in\{3, 4\}$ and $r\in\{8, 10\}$. Some observations can be made from the figure: when both $m$ and $r$ are sufficiently small, the minimal spreading allocation is optimal for both $P_s(\alpha)$ and $\mu_{s}(\alpha)$. When both $m$ and $r$ are sufficiently large, the maximal spreading allocation is optimal for both performance metrics. Otherwise, we cannot find an optimal allocation to simultaneously optimize both performance metrics. For example, when $m=4$ and $r=8$, $P_s(\alpha)$ reaches its maximum at $\alpha=1$, while  $\mu_{s}(\alpha)$ reaches its maximum at $\alpha=4$. A performance tradeoff can be achieved by choosing $\alpha=2$ to obtain acceptable $P_s(\alpha)$ and $\mu_{s}(\alpha)$.

\subsection{Probabilistic Access Model}
For probabilistic access model under scaled exponential service time, the DSS service rate \eqref{eq:service} becomes
\begin{equation}
      \mu_s(\alpha)=\sum^{m\alpha}_{\varphi=\alpha}\frac{\mu\alpha}{H_\varphi-H_{\varphi-\alpha}} \binom {m\alpha}{\varphi} (1-p)^{\varphi}p^{m\alpha-\varphi}.
        \label{Eq:prob_scale}
\end{equation}
By comparing \eqref{eq:service_prob_exp} and \eqref{Eq:prob_scale}, we expect that the minimal spreading allocation may not be always optimal. To this end, we first present the following result on the optimal $\alpha$ to maximize the service rate when no redundancy is used.

\begin{theorem}[optimal $\alpha$]
\label{Le:prob_scale}
Under the probabilistic access model with a scaled exponential service time and considering a no redundancy scenario, i.e. $m=1$, the optimal $\alpha$ which maximizes the DSS service rate $\mu_s(\alpha)$ is located in the range $[(1/2-p)/p,(1-p)/p]$.
\end{theorem}
\begin{IEEEproof}
Since the value of $\alpha$ is an integer, $\mu_s(\alpha)$ as a function of $\alpha$ is discrete. To prove the optimal $\alpha$ is located in the range $[(1/2-p)/p,(1-p)/p]$, we need to show $\mu_s(\alpha)$ increases with $\alpha$ when $\alpha\le (1/2-p)/p$ (i.e. $\mu_s(\alpha)\le\mu_s(\alpha+1)$) and decreases with increasing $\alpha$ when $\alpha\ge (1-p)/p$ (i.e. $\mu_s(\alpha)\ge\mu_s(\alpha+1)$).
For $\alpha\ge 1$, the DSS service rate is $\mu_s(\alpha)= \frac{\mu \alpha }{H_{\alpha}}(1-p)^{\alpha}$, 
resulting in $\mu_s(\alpha)/\mu_s(\alpha+1)= (\alpha H_{\alpha+1} )/((\alpha+1) H_{\alpha}(1-p))$.
Thus,
\begin{equation}
\begin{split}
\frac{\mu_s(\alpha)}{\mu_s(\alpha+1)}\ge1 &\Leftrightarrow \frac{ \alpha H_{\alpha+1} }{(\alpha+1) H_{\alpha}(1-p)}\ge 1\\
&\Leftrightarrow \frac{\alpha}{\alpha+1}\ge (1-\alpha p -p)H_{\alpha}.
\end{split}
\label{Eq:prob_scale_sp1}
\end{equation}
Since $\frac{\alpha}{\alpha+1}>0$, \eqref{Eq:prob_scale_sp1} is satisfied when $(1-\alpha p -p)H_{\alpha}\le0$, resulting in $\alpha \ge (1-p)/p$.
Similarly,
\begin{equation}
\begin{split}
\frac{\mu_s(\alpha)}{\mu_s(\alpha+1)}\le 1 \Leftrightarrow \frac{\alpha}{\alpha+1}\le (1-\alpha p -p)H_{\alpha}.
\end{split}
\label{Eq:prob_scale_sp2}
\end{equation}
On the other hand, since { $\frac{1}{2}H_{\alpha}- \frac{\alpha}{\alpha+1}=  \frac{1}{2} (H_{\alpha+1}+ \frac{1}{\alpha+1}) -1=\frac{1}{2} (\sum_{i=2}^{\alpha+1} \frac{1}{i} + \frac{1}{\alpha+1}) -\frac{1}{2} 
 \ge\frac{1}{2} (\frac{\alpha}{\alpha+1}+ \frac{1}{\alpha+1}) -\frac{1}{2}=0$}, we have { $\frac{\alpha}{\alpha+1}\le 1/2H_{\alpha}$}.
Thus, if $(1-\alpha p -p)\ge 1/2$, or equivalently $\alpha \le (1/2-p)/p$, \eqref{Eq:prob_scale_sp2} holds.
\end{IEEEproof}

From Theorem~\ref{Le:prob_scale}, it is easy to see that $p$ is small, the optimal $\alpha$ is greater than $1$. Then we have the following remark on the minimal spreading allocation.
\begin{remark}
 Under the probabilistic access model with a scaled exponential service time, the minimal spreading allocation ($\alpha=1$) does not always maximize the service rate.
 \label{Th:prob_scale}
\end{remark}

\subsubsection{Optimality and Non-optimality Conditions for Minimal Spreading Allocation}
From Remark~\ref{Th:prob_scale}, we know that the minimal spreading allocation is not always optimal. Considering the complexity of \eqref{Eq:prob_scale}, finding the $\alpha$ that maximizes $\mu_s(\alpha)$ is hard. Similar to the fixed-size access model in Sec.~\ref{subsubsec:condition1}, we find  the optimality and non-optimality conditions  for  the minimal  spreading  allocation.  

\begin{theorem}[minimal spreading optimality condition]
 Under the probabilistic access model with a scaled exponential service time, the minimal spreading allocation maximizes the DSS service rate $\mu_s(\alpha)$ when\\ { $p\ge\max_{\alpha\ge2}\{1-\frac{1}{\sqrt[\alpha-1]{\alpha\binom{m\alpha-1}{\alpha-1}}}\}$}.
 \label{Th:prob_opt_scale}
\end{theorem}
\begin{IEEEproof}
To prove the theorem statement, we need to find the condition ensuring $\mu_s(1)\ge \mu_s(\alpha)$ for all $\alpha\ge2$. Using \eqref{Eq:prob_scale} and considering that $\mu_s(\alpha|\varphi)<\mu\varphi$ for $\alpha\ge2$ according to \eqref{ieq:rate}, we have
\begin{align*}
    \mu_s(\alpha)&<\mu\sum^{m\alpha}_{\varphi=\alpha}\varphi\binom{m\alpha}{\varphi}(1-p)^{\varphi}p^{m\alpha-\varphi}\\
    &=\mu m\alpha\sum^{m\alpha}_{\varphi=\alpha}(\prod^{\alpha-2}_{i=0}\frac{m\alpha-1-i}{\varphi-1-i})\binom{m\alpha-\alpha}{\varphi-\alpha}(1-p)^{\varphi}p^{m\alpha-\varphi}.
     \end{align*}
Since $\varphi$ goes from $\alpha$ to $m\alpha$, we have
    \begin{align*}
    \mu_s(\alpha)&<\mu m\alpha\binom{m\alpha-1}{\alpha-1}(1-p)^{\alpha}\cdot\\ &\sum^{m\alpha-\alpha}_{\varphi=0}\binom{m\alpha-\alpha}{\varphi}(1-p)^{\varphi}p^{m\alpha-\alpha-\varphi}\\
    &=\mu m\alpha\binom{m\alpha-1}{\alpha-1}(1-p)^{\alpha} ~~~~
    \text{(by binomial\ expansion)}.
\end{align*}
From \eqref{Eq:prob_scale}, $\mu_s(1)=\mu\sum^{m}_{\varphi=1}\varphi\binom{m}{\varphi}(1-p)^{\varphi}p^{m-\varphi}
    =\mu m(1-p)$.
Now, to satisfy $\mu_s(\alpha)\le\mu_s(1)$, we have $\mu m\alpha\binom{m\alpha-1}{\alpha-1}(1-p)^{\alpha}\le\mu m(1-p)
   \Leftrightarrow (1-p)^{\alpha-1}\le\frac{1}{\alpha\binom{m\alpha-1}{\alpha-1}}.$
\noindent Thus, if { $p\ge1-\frac{1}{\sqrt[\alpha-1]{\alpha\binom{m\alpha-1}{\alpha-1}}}$} holds for all { $\alpha \ge 2$}, $\mu_s(1)$ is optimal.
\end{IEEEproof}



\begin{theorem}[minimal spreading non-optimality condition]
Under the probabilistic access model with a scaled exponential service time, the minimal spreading allocation  does not maximize the DSS service rate $\mu_s(\alpha)$ when { $p\le \max_{\alpha\ge2 }\{1-\sqrt[\alpha-1]{\frac{m}{m\alpha-\alpha+1}}\}$}.
 \label{Th:prob_nonopt_scale}
\end{theorem}

\begin{IEEEproof}
We are interested in finding a condition that guarantees the existence of an $\alpha \ge 2$, such that $\mu_s(1)\le \mu_s(\alpha)$. The expression for $\mu_s(\alpha)$ is given in \eqref{Eq:prob_scale}. According to \eqref{ieq:rate}, we have $\mu_s(\alpha|\varphi)>\mu(\varphi-\alpha+1)$ when $\alpha\ge2$, then
\begin{align*}
    &\mu_s(\alpha)>\mu\sum^{m\alpha}_{\varphi=\alpha}(\varphi-\alpha+1)\binom{m\alpha}{\varphi}(1-p)^{\varphi}p^{m\alpha-\varphi}\\
    &=\mu\sum^{m\alpha}_{\varphi=\alpha}(\varphi-\alpha+1)\binom{m\alpha-\alpha+1}{\varphi-\alpha+1}(1-p)^{\varphi}p^{m\alpha-\varphi} \prod^{\alpha-2}_{i=0}\frac{m\alpha-i}{\varphi-i}   \end{align*}
Since $\varphi$ goes from $\alpha$ to $m\alpha$, thus
    \begin{align*}
    &\mu_s(\alpha)>\mu\sum^{m\alpha}_{\varphi=\alpha}(\varphi-\alpha+1)\binom{m\alpha-\alpha+1}{\varphi-\alpha+1}(1-p)^{\varphi}p^{m\alpha-\varphi}\\
    &=\mu(m\alpha-\alpha+1)(1-p)^{\alpha}\sum^{m\alpha-\alpha}_{\varphi=0}\binom{m\alpha-\alpha}{\varphi}(1-p)^{\varphi}p^{m\alpha-\alpha-\varphi}\\
    &=\mu(m\alpha-\alpha+1)(1-p)^{\alpha}.
\end{align*}
Since $\mu_s(1)=\mu m(1-p)$, to satisfy $\mu_s(\alpha)\ge\mu_s(1)$, it is sufficient to have $\mu(m\alpha-\alpha+1)(1-p)^{\alpha}\ge\mu m(1-p)
   \Leftrightarrow (1-p)^{\alpha-1}\ge\frac{m}{m\alpha-\alpha+1}$.
\noindent Thus, if $p\le 1-\sqrt[\alpha-1]{\frac{m}{m\alpha-\alpha+1}}$ holds for an $\alpha \ge 2$, $\mu_s(1)$ is not optimal.
\end{IEEEproof}


From Theorems~\ref{Th:prob_opt_scale} and \ref{Th:prob_nonopt_scale}, we see that both the optimality and non-optimality conditions for the minimal spreading allocation depend on the probability of failed access $p$. Roughly speaking, when $p$ is close to $1$, then $\alpha=1$ is optimal while for $p$ close to $0$, an $\alpha>1$ is optimal. 
Based on these theorems, we conjecture the following about $\alpha$ that maximizes $\mu_s(\alpha)$. 
\begin{conjecture}
The following assertions are true:
\begin{enumerate}
\item Under the probabilistic access model, given the redundancy level $m$, the optimal $\alpha$ increases as $p$ decreases. 
\item For every $m$, there exists a $\gamma\in(0,1)$, such that for all $p\ge\gamma$, the minimal spreading is optimal. 
\item For every $m$, there exists a $\zeta\in(0,1)$, such that for all $p\le\zeta$, the maximal spreading is optimal. 
\end{enumerate}
\label{hyp:prob1}
\end{conjecture}
\begin{remark}
The derived bounds on $p$ that guarantee the optimality or non-optimality of minimal spreading are not tight. Thus, there exists a gap between these bounds. Consider, for example, the probabilistic access model and $m=2$. By applying the conditions in Theorem~\ref{Th:prob_opt_scale} and \ref{Th:prob_nonopt_scale}, we arrive at an optimality sufficient condition of $p\ge 0.83$ and a non-optimality sufficient condition of $p\le 0.33$. 
\label{rem:tight}
\end{remark}

\subsubsection{Numerical Analysis}
\begin{figure}[hbt]
	\centering
	\includegraphics[width=0.725\textwidth]{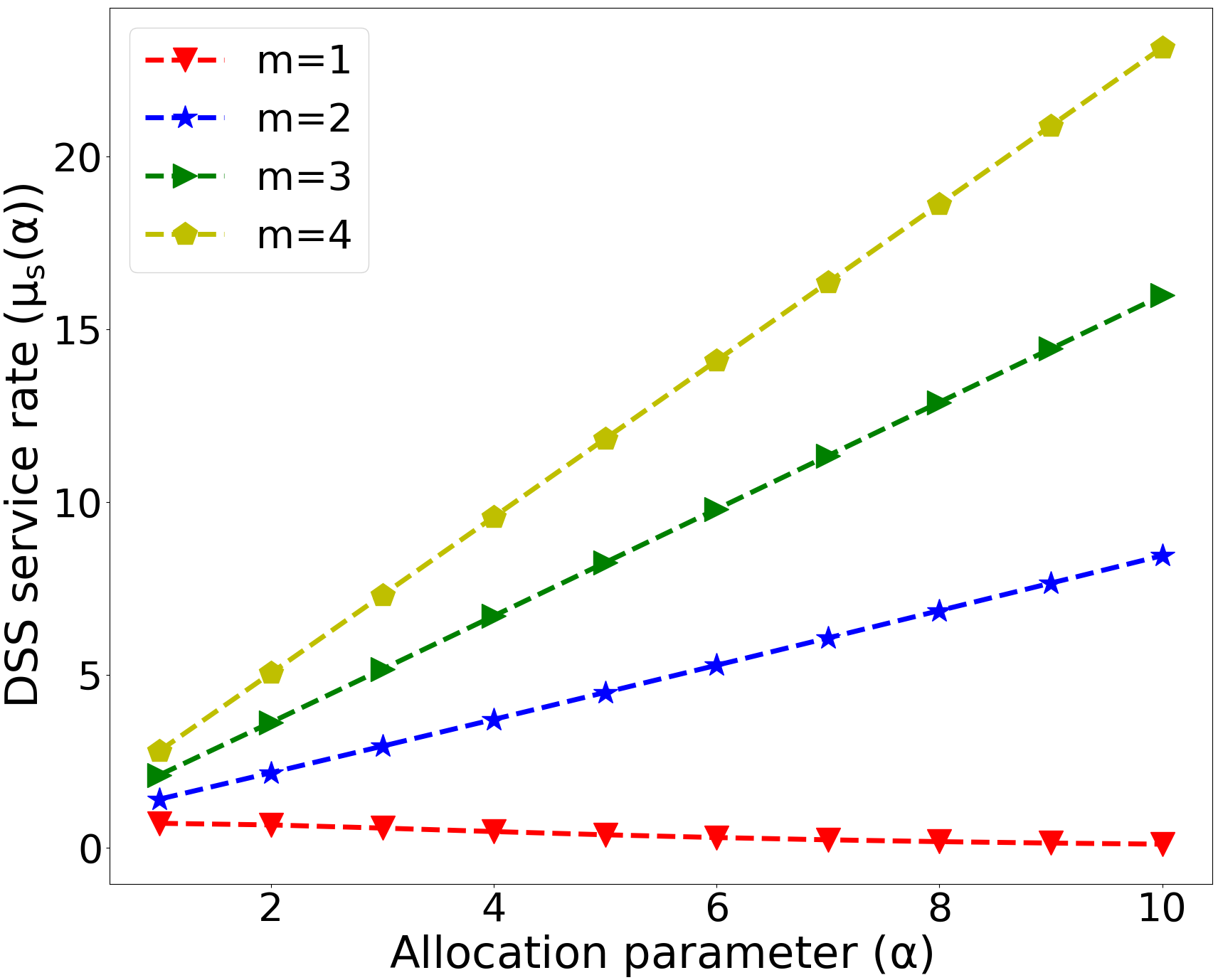}~\\
	\includegraphics[width=0.725\textwidth]{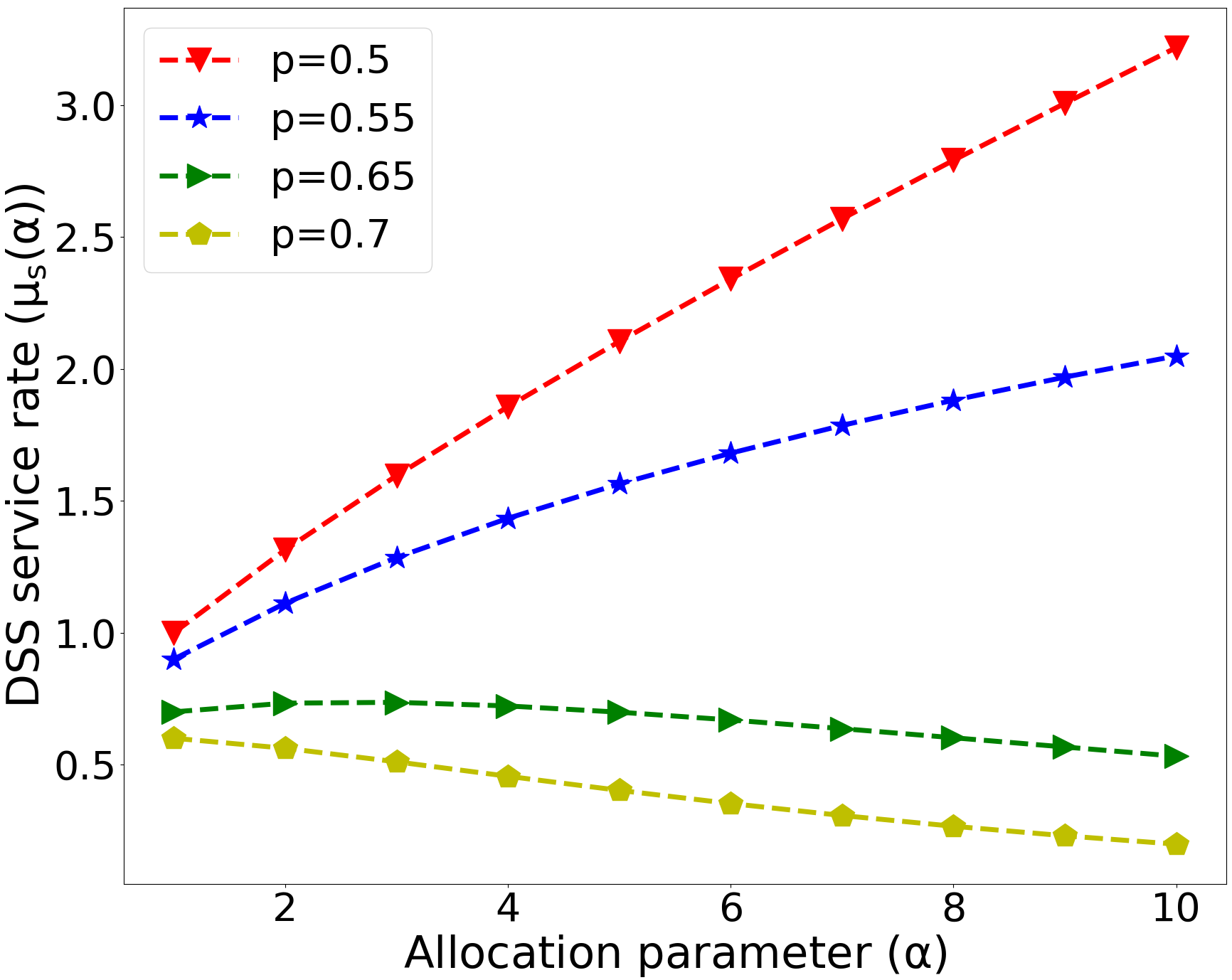}~
	\caption{ The DSS service rate $\mu_{s}(\alpha)$ for the probabilistic access model with scaled exponential service time as a function of the allocation parameter $\alpha$ (cf.~\eqref{Eq:prob_scale}).  The number of storage nodes is $N\ge m\alpha$, and the service time follows $\expD(1/\alpha)$. (upper) $\mu_{s}(\alpha)$ vs. $\alpha$ with the failed access probability $p=0.3$ for four values of $m$. (lower) $\mu_{s}(\alpha)$ vs. $\alpha$ with redundancy of $m=2$ for four values of $p$. Given $p$ (or $m$), the optimal allocation changes from the minimal spreading allocation to the maximal spreading allocation as $m$ increases or $p$ decreases.}
	\label{fig:large_scale_prob_mp}
\end{figure}

In Fig.~\ref{fig:large_scale_prob_mp}, we evaluate the expression in \eqref{Eq:prob_scale} for $\mu_{s}(\alpha)$ to see how the DSS service rate changes with  $\alpha$. We consider a system with $N\ge m\alpha$ storage nodes.
Using Theorems~\ref{Th:prob_opt_scale} and \ref{Th:prob_nonopt_scale}, we can easily calculate the optimality and non-optimality conditions. For example, when $m=2$, the minimal spreading allocation is optimal when $p\ge 0.83$ and is non-optimal when $p\le 0.3$. These conditions provide some knowledge of the optimal allocation and further insight on the optimal allocation can be found from Fig.~\ref{fig:large_scale_prob_mp}. 
The upper graph shows $\mu_{s}(\alpha)$ vs. $\alpha$ for four different levels of redundancy $m\in\{1, 2, 3, 4\}$, and the failed access probability $p=0.3$. The lower graph shows $\mu_{s}(\alpha)$ vs. $\alpha$ for four different values of $p\in\{0.5, 0.55, 0.65, 0.7\}$, and the redundancy level  $m=2$.
In the upper subfigure, when $m=1$, $\mu_{s}(\alpha)$ decreases with increasing $\alpha$ and reaches its maximum at $\alpha=1$, i.e. the minimal spreading allocation is optimal. When $m\ge2$, $\mu_{s}(\alpha)$ increases with $\alpha$ and reaches its maximum at $\alpha=10$, i.e. the maximal spreading allocation is optimal. In the lower subfigure, when $p\le 0.55$, the maximal spreading allocation is optimal. When $p=0.65$, $\alpha=2$ allocation is optimal. When $p=0.7$, the minimal spreading allocation is optimal. Therefore, the optimal allocation changes with $p$.

\begin{figure}[hbt]
\centering
\includegraphics[width=0.725\textwidth]{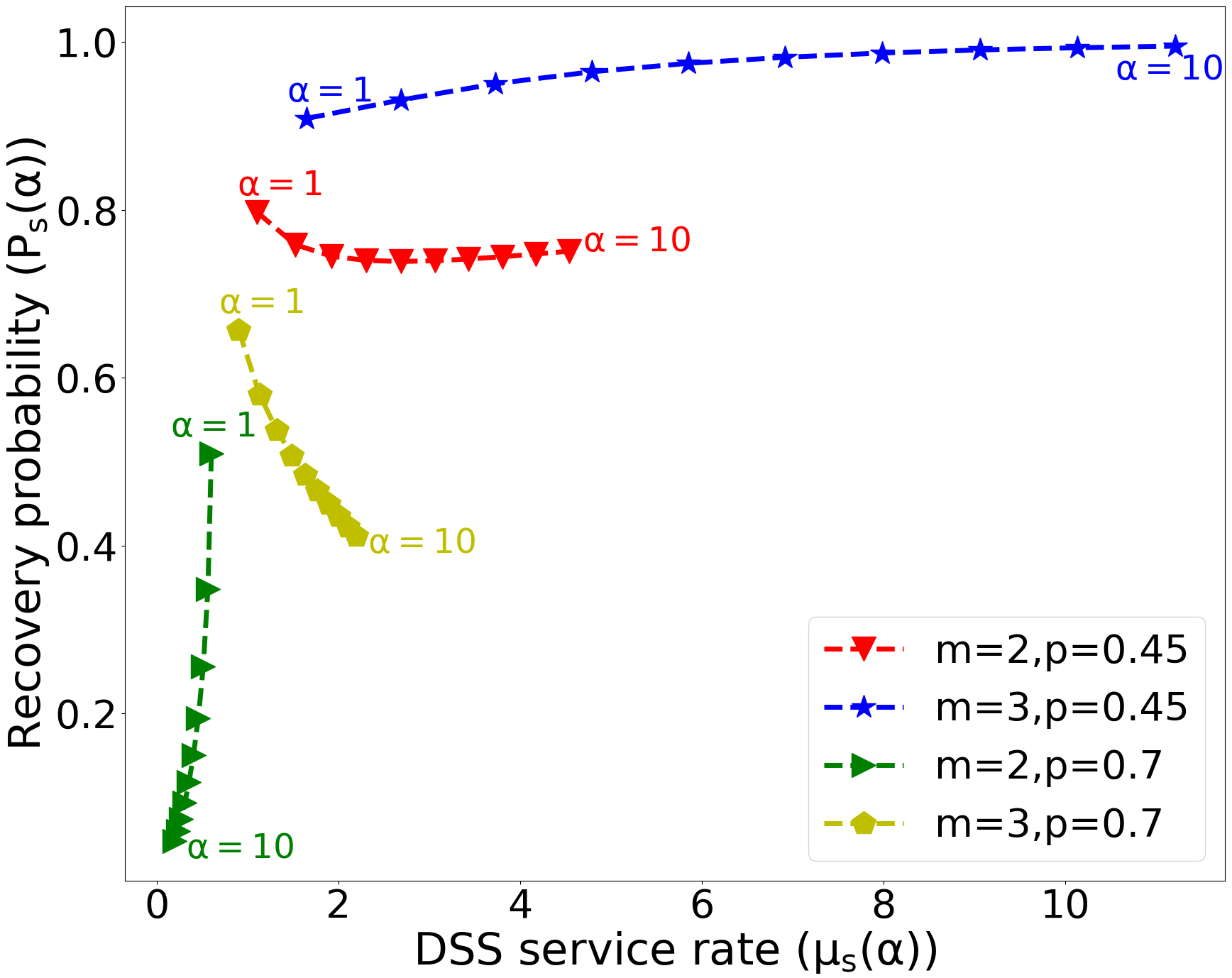}
	\caption{ The successful recovery probability $P_s(\alpha)$ vs. the DSS service rate $\mu_{s}(\alpha)$ for the probabilistic access model as a function of $\alpha$ for different values of $m$ and $r$ (cf.~\eqref{eq:prob_prob} and \eqref{Eq:prob_scale}). The number of storage nodes is $N\ge m\alpha$, and the service time follows $\expD(1/\alpha)$. The optimal allocation is affected by both $m$ and $r$ values. \label{fig:large_scale_prob_up}}
\end{figure}

In Fig.~\ref{fig:large_scale_prob_up}, we analyze $P_s(\alpha)$ vs. $\mu_{s}(\alpha)$ as $\alpha$ increases from $1$ to $10$. We consider a system with $N\ge m\alpha$ storage nodes and two values for each parameter $m\in\{2, 3\}$ and $p\in\{0.45, 0.7\}$. When $m$ is sufficiently small and $p$ is sufficiently large, the minimal spreading allocation is optimal for both $P_s(\alpha)$ and $\mu_{s}(\alpha)$, while for sufficiently large $m$ and sufficiently small $p$, the maximal spreading allocation is optimal for both performance metrics. For a general scenario, the optimal $\alpha$ for maximizing the service rate may be different from that for maximizing the probability of recovery. For example, when $m=3$ and $p=0.7$, $P_s(\alpha)$ reaches its maximum at $\alpha=1$, and  $\mu_{s}(\alpha)$ reaches its maximum at $\alpha=10$.

\section{Storage Allocation for Large Files with Shifted Exponential service Time}
 \label{sec:large_shift}
Here the service time follows a shifted exponential distribution with the shift $\Delta/\alpha$ and the rate $\mu$, i.e., $\sexpD(\Delta/\alpha,\mu)$. We determine the $\mu_{s}(\alpha)$ for the two considered access models. Some of the results in this section were published in \cite{peng2018distributed}.

 \subsection{Fixed-Size Access Model}
 For fixed-size access model under shifted exponential service time, the DSS service rate \eqref{eq:service} becomes
\begin{equation}
      \mu_s(\alpha)=\frac{\mu\alpha}{\binom N r}\!\! \sum^{\min(r,m\alpha)}_{\varphi=\alpha}\!\!
      \frac{
      \binom {m\alpha}{\varphi} \binom {N-m\alpha}{r-\varphi}}
      {\Delta\mu+\alpha(H_{\varphi}-H_{\varphi-\alpha})} 
        \label{eq:service_shift}
\end{equation}
According to \eqref{eq:service_shift}, the minimal spreading allocation may not be always optimal. In fact, we find a special scenario where there exists an optimal $\alpha\ge2$ that maximizes the service rate.

Let us start by assuming a constant service time of $\Delta$ for $k$ block. In this case,  { $\mu_s(\alpha|\varphi(\mathcal{A}))=\alpha/\Delta$} for a set $\mathcal{A}$.
Therefore, for the maximal spreading allocation, we have $\mu_s(r)= \frac{ r\binom {r m}{r} }{\Delta\binom{N}{r}}$.
For the minimal spreading allocation, i.e. $\alpha=1$, using Vandermonde's convolution, we arrive at { $\mu_s(1)= \frac{ 1}{\Delta} (1-\frac{\binom{N-m}{r}}{\binom{N}{r}})$}.

\begin{theorem}[minimal vs. maximal spreading]
\label{Le:fixed_spreading_shift}
Under the fixed-size access model and a constant service time $\Delta$, the maximal spreading allocation outperforms the minimal spreading allocation when $rm\ge N$.
\end{theorem}
\begin{IEEEproof}
Since $rm\ge N$ and $\binom{N-m}{r}\ge 0$, { $\mu_s(r) = \frac{ r\binom {r m}{r} }{\Delta\binom{N}{r}} \ge \frac{ \binom {r m}{r}-\binom{N-m}{r} }{\Delta\binom{N}{r}}\ge \frac{ \binom {N}{r}-\binom{N-m}{r} }{\Delta\binom{N}{r}}=\mu_s(1)$}.
\end{IEEEproof}


\begin{remark}
 Under the fixed-size access model with a constant service time $\Delta$, the minimal spreading allocation ($\alpha=1$) does not always maximize the service rate.
 \label{Th:shift-fix}
\end{remark}

The constant service time is a special case of the shifted exponential service time.  We next find the optimality and non-optimality conditions for the minimal spreading allocation under shifted exponential service time. Our findings show that Remark~\ref{Th:shift-fix} is also true for the shifted exponential service time.

\subsubsection{Optimality and Non-optimality Conditions for the Minimal Spreading Allocation}
finding the optimal $\alpha$ that maximizes $\mu_s(\alpha)$ in \eqref{eq:service_shift} is difficult. The proofs of the following theorems are similar to the proofs of Theorem~\ref{Th:fix_opt_scale} and Theorem~\ref{Th:fix_nonopt_scale}; see the supplementary Appendix.

\begin{theorem}[minimal spreading optimality condition]
 Under the fixed-size access model a shifted exponential service time, the minimal spreading allocation maximizes the DSS service rate $\mu_s(\alpha)$ when \\{$ r\le \min_{2\le \alpha\le r} \{ 1+\sqrt[\alpha-1]{\frac{\Delta\mu+\alpha}{\alpha(\Delta\mu m+1) \binom{m\alpha-1}{\alpha-1}}} (N-1)\}$}.
 \label{Th:fix_opt_shift}
\end{theorem}


\begin{theorem}[minimal spreading non-optimality condition]
Under the fixed-size access model and a shifted exponential service time, the minimal spreading allocation does not maximize the DSS service rate $\mu_s(\alpha)$ when\\ {$r\ge\min_{2\le\alpha\le r} \{\sqrt[\alpha-1]{ \frac{\Delta\mu m(m\alpha-\alpha+1)+m\alpha^2 }{\alpha(\Delta\mu+1)(m\alpha-\alpha+1)}}\cdot  (N-\alpha+1) +\alpha-1\}$}.
 \label{Th:fix_nonopt_shift}
\end{theorem}


The conditions in Theorems~\ref{Th:fix_opt_shift} and \ref{Th:fix_nonopt_shift} depend on the number of accessed nodes $r$. For the shifted exponential service, we have Conjecture~\ref{hyp:fix1}, providing guidelines on finding the optimal $\alpha$ for maximizing $\mu_s(\alpha)$. 

\subsubsection{Numerical Analysis}
\begin{figure}[hbt]
	\centering
	\includegraphics[width=0.725\textwidth]{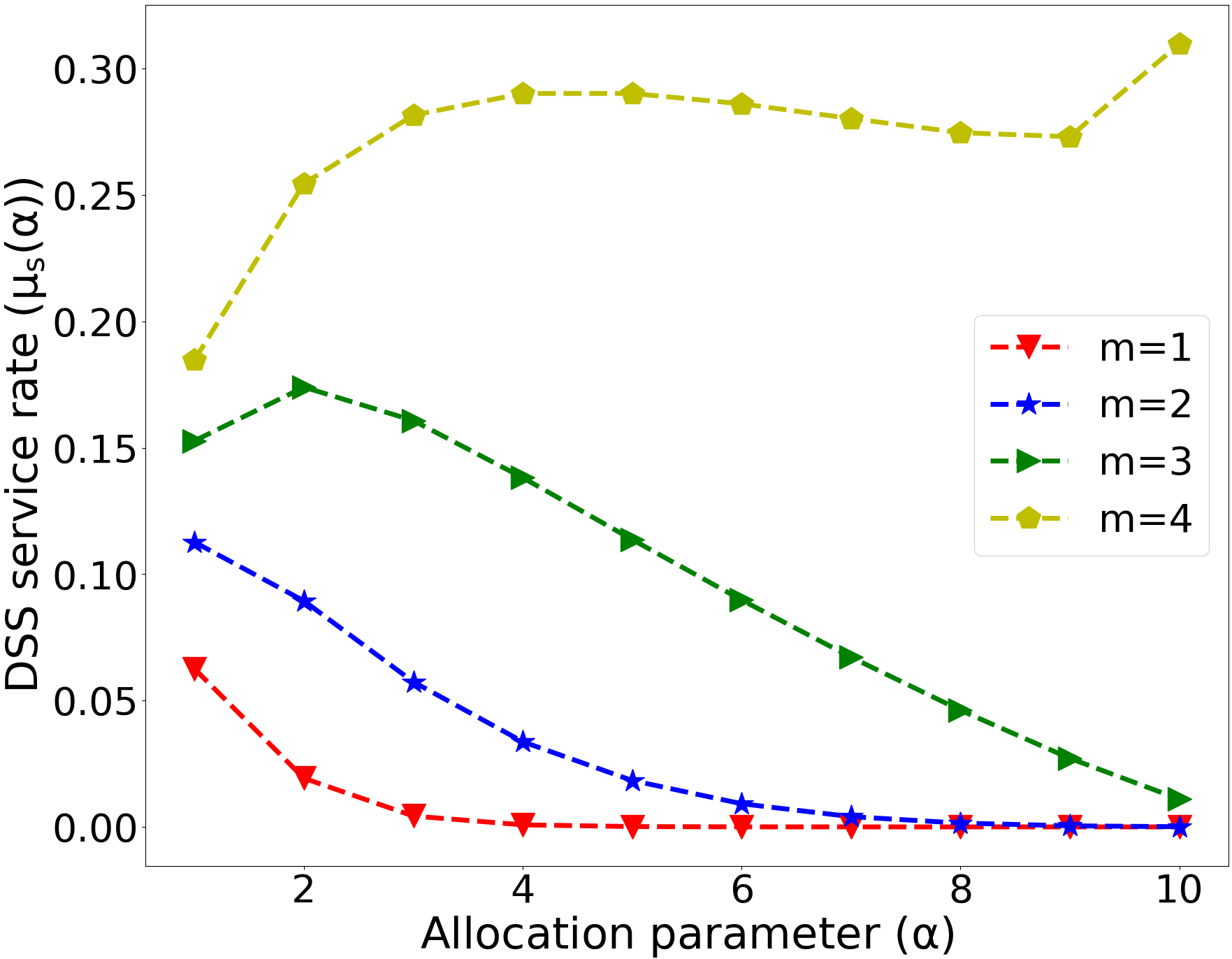}~\\
	\includegraphics[width=0.725\textwidth]{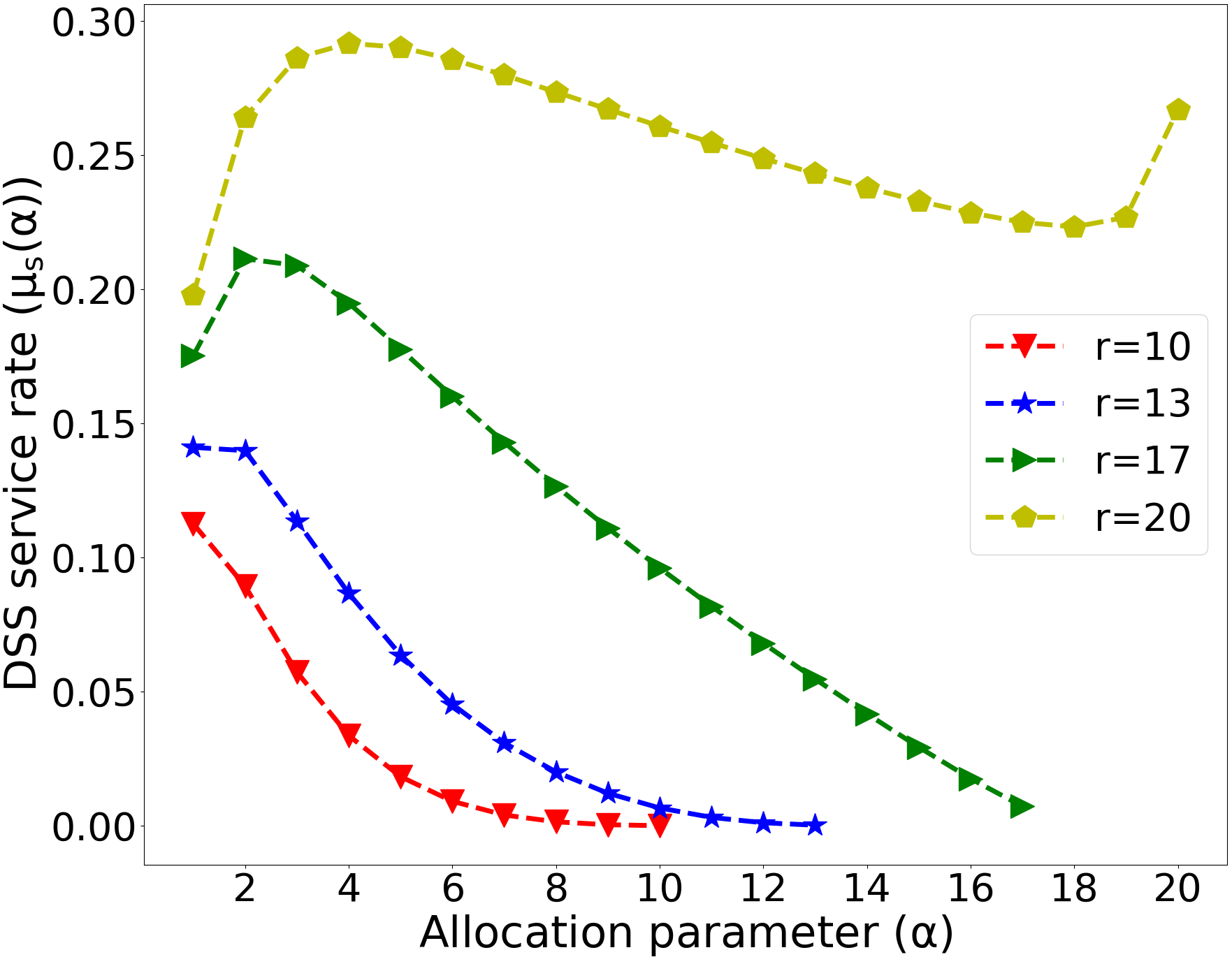}~
	\caption{Service rate $\mu_{s}(\alpha)$ for the fixed-size access model with shifted exponential service time as a function of the allocation parameter $\alpha$ (cf.~\eqref{eq:service_shift}).  The number of storage nodes is $N=40$, and the service time follows $\sexpD(3,1)$. (upper) $\mu_{s}(\alpha)$ vs. $\alpha$ with $r=10$ accessed nodes for four values of $m$. (lower) $\mu_{s}(\alpha)$ vs. $\alpha$ with $m=2$ redundancy for four values of $r$. Given $r$ (or $m$), the optimal allocation changes from the minimal spreading allocation to the maximal spreading allocation as $m$ (or $r$) increases.}
	\label{fig:large_shift_fix_mr}
\end{figure}

In Fig.~\ref{fig:large_shift_fix_mr}, we evaluate the expression \eqref{eq:service_shift}  for $\mu_{s}(\alpha)$ to see how the DSS service rate changes with  $\alpha$. We consider a system with $N=40$ storage nodes. 
Using Theorems~\ref{Th:fix_opt_shift} and \ref{Th:fix_nonopt_shift}, we can easily calculate the optimality and non-optimality conditions. For example, when $m=2$, the minimal spreading allocation is optimal when $r<6$ and is non-optimal when $r> 36$. These conditions only provide limited knowledge of the optimal allocation and further insight on the optimal allocation can be found from Fig.~\ref{fig:large_shift_fix_mr}.
The upper graph shows $\mu_{s}(\alpha)$ vs. $\alpha$ for four different redundancy levels $m\in\{1, 2, 3, 4\}$ where the number of accessed nodes is $r=10$. The lower graph shows $\mu_{s}(\alpha)$ vs. $\alpha$ for $r\in\{10, 13, 17, 20\}$, and the redundancy level $m=2$. 
In the upper subfigure, when $m\le2$, the minimal spreading allocation is optimal. When $m=3$, $\mu_{s}(\alpha)$ reaches its maximum at $\alpha=3$. When $m=4$, the maximal spreading allocation is optimal. In the lower subfigure, when $r\le13$, the minimal spreading allocation is optimal, while for $r=17$, the allocation with $\alpha=2$ is optimal. When $r=20$, although the allocation with $\alpha=4$ is optimal, the maximal spreading allocation provides a local maximum value which is close to the global maximum value. From the observations, we conclude that the minimal spreading allocation is optimal only when $m$ or $r$ is sufficiently small. Otherwise, an allocation with $\alpha\ge2$ is optimal.

\begin{figure}[hbt]
\centering
\includegraphics[width=0.725\textwidth]{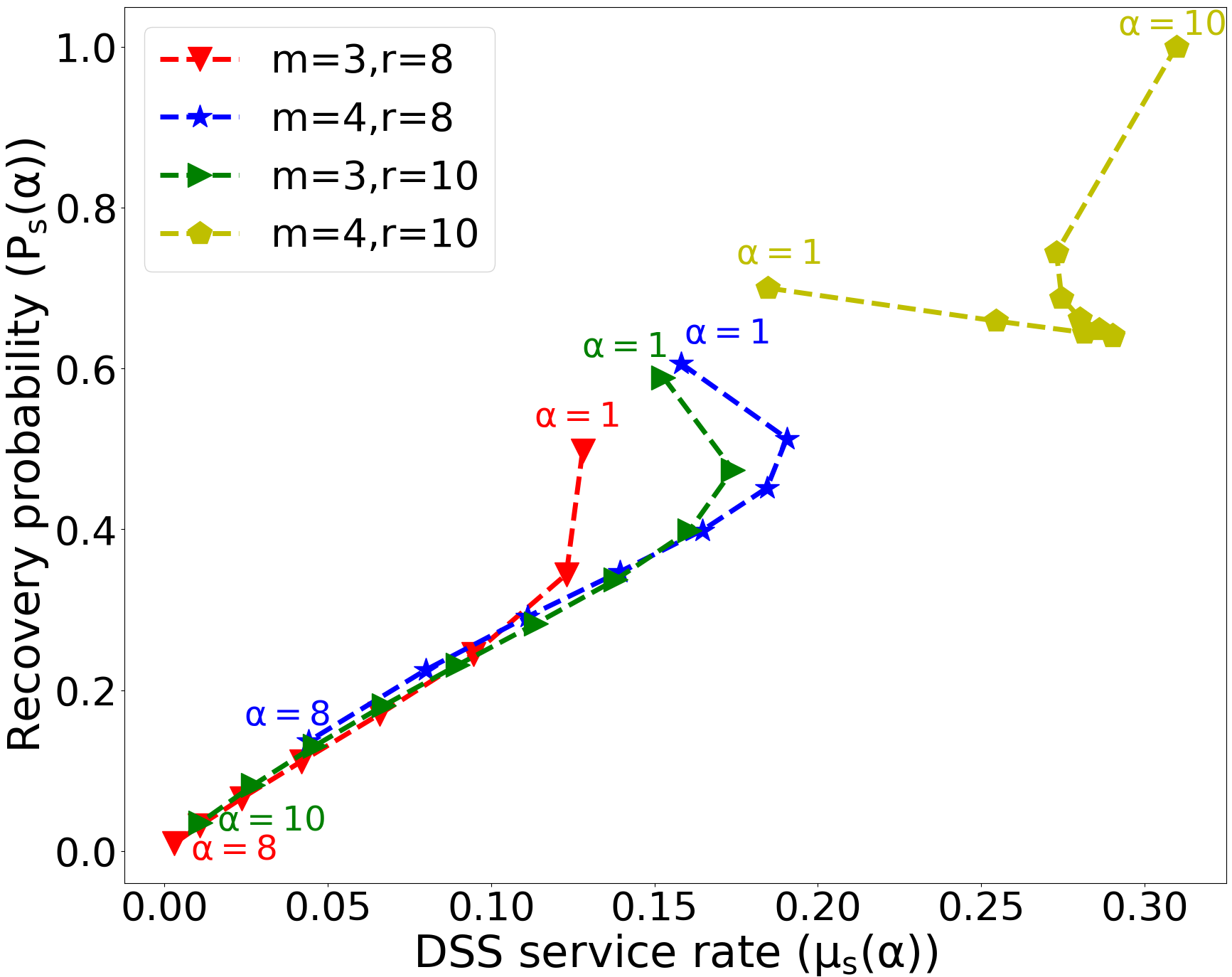}
	\caption{ Successful recovery probability $P_s(\alpha)$ vs. the DSS service rate $\mu_{s}(\alpha)$ for the fixed-size access model as a function of $\alpha$ for different values of $m$ and $r$ (cf.~\eqref{eq:prob_fix} and \eqref{eq:service_shift}). The number of storage nodes is $N=40$, and the service time follows $\sexpD(3,1)$. The minimal (or maximal) spreading allocation is optimal when $m$ and $r$ are sufficiently small (or large). Otherwise, the optimal allocation is different for different performance metrics. \label{fig:large_shift_fix_up}  }
\end{figure}

In Fig.~\ref{fig:large_shift_fix_up}, we analyze $P_s(\alpha)$ vs. $\mu_{s}(\alpha)$ as $\alpha$ increases from $1$ to $r$. We consider a system with $N=40$ storage nodes, $m\in\{3, 4\}$, and $r\in\{8, 10\}$. Some observations can be made from the figure: when both $m$ and $r$ are sufficiently small, e.g. $m=3$ and $r=8$, the minimal spreading allocation is optimal for both $P_s(\alpha)$ and $\mu_{s}(\alpha)$. When both $m$ and $r$ are sufficiently large, e.g. $m=4$ and $r=10$, the maximal spreading allocation is optimal. Otherwise, e.g. $m=3$ and $r=10$, there is no optimal $\alpha$ that maximizes both $P_s(\alpha)$ and $\mu_{s}(\alpha)$ simultaneously. 

\subsection{Probabilistic Access Model}
For the probabilistic access and a shifted exponential service, service rate \eqref{eq:service} becomes
\begin{equation}
      \mu_s(\alpha)=\sum^{m\alpha}_{\varphi=\alpha}
      \binom {m\alpha}{\varphi}
      \frac{\mu\alpha\cdot (1-p)^{\varphi}p^{m\alpha-\varphi}}{\Delta\mu+\alpha(H_{\varphi}-H_{\varphi-\alpha})} .
        \label{Eq:prob_shift}
\end{equation}
From \eqref{Eq:prob_shift}, one can expect that the minimal spreading allocation may not be always optimal. To this end, we present the following result on the optimal $\alpha$ to maximize the service rate when no redundancy is used to store the data.

Assume a constant service time of $\Delta$ for $k$ block. In this case,  $\mu_s(\alpha|\varphi(\mathcal{A}))=\alpha/\Delta$ for a set $\mathcal{A}$. Therefore, for no redundancy scenario, i.e. $m=1$, we have $\mu_s(\alpha)= \frac{\alpha }{\Delta}(1-p)^{\alpha}$.
\begin{theorem}[optimal $\alpha$]
\label{Le:prob_shift}
Under the probabilistic access model, a constant service time $\Delta$, and no redundancy case in data storage ($m=1$), the DSS service rate $\mu_s(\alpha)$ reaches its maximum when  $\alpha=\lceil p/(1-p)\rceil$ or $\alpha=\lfloor p/(1-p)\rfloor$.
\end{theorem}
\begin{IEEEproof}
Given an integer $\alpha\ge 1$, the DSS service rate
$\mu_s(\alpha)= \frac{ \alpha }{\Delta} (1-p)^{\alpha}$ is discrete.
Then we find the optimal $\alpha$ by comparing the ratio $\mu_s(\alpha)/\mu_s(\alpha+1)=  \alpha/(\alpha+1 (1-p))$ with $1$.
Thus,
{ $\frac{\mu_s(\alpha)}{\mu_s(\alpha+1)}\ge1 \Leftrightarrow \frac{ \alpha  }{\alpha+1 (1-p)}\ge 1
\Leftrightarrow \alpha\le \frac{p}{1-p}$}.
Similarly,
{ $\frac{\mu_s(\alpha)}{\mu_s(\alpha+1)}\le 1 \Leftrightarrow\alpha\ge  \frac{p}{1-p}$}.
Since $\alpha$ is an integer, $\mu_s(\alpha)$ reaches the maximum at $\lceil\frac{p}{1-p} \rceil$ or  $\lfloor\frac{p}{1-p} \rfloor$.
\end{IEEEproof}
\begin{remark}
Under the probabilistic access model with a constant service time $\Delta$, the minimal spreading allocation ($\alpha=1$) does not always maximize the DSS service rate.
 \label{Th:shift-prob}
\end{remark}
Now, we go one step further and find the optimality and non-optimality conditions for the minimal spreading allocation when service time follows a shifted exponential distribution.

\subsubsection{Optimality and Non-optimality Conditions for the Minimal Spreading Allocation}
Considering the complexity of \eqref{Eq:prob_shift}, finding the optimal $\alpha$ that maximizes $\mu_s(\alpha)$ in a general scenario is difficult.
\begin{theorem}[minimal spreading optimality condition]
 Under the probabilistic access model with a shifted exponential service time, the minimal spreading allocation maximizes the DSS service rate $\mu_s(\alpha)$ when\\ { $ p\ge\max_{\alpha\ge2}\Bigl\{1 - \sqrt[\alpha-1]{(\Delta\mu+\alpha)/{\alpha(\Delta\mu m+1)\binom{m\alpha-1}{\alpha-1}}}\Bigr\}$}.
 \label{Th:prob_opt_shift}
\end{theorem}
\begin{IEEEproof}
Similar to the proof in Theorem~\ref{Th:prob_opt_scale}. For details, please refer to the supplementary  Appendix. 
\end{IEEEproof}

\begin{theorem}[minimal spreading non-optimality condition]
 Under the probabilistic access model with a scaled exponential service time, the minimal spreading allocation does not maximize the DSS service rate $\mu_s(\alpha)$  when\\ { $p\le \max_{\alpha\ge2}\Bigl\{1 - \sqrt[\alpha-1]{\frac{m(\Delta\mu(m\alpha-\alpha+1)+\alpha^2)}{\alpha(\Delta\mu +1)(m\alpha-\alpha+1)}}\Bigr\}$}.
 \label{Th:prob_nonopt_shift}
\end{theorem}
\begin{IEEEproof}
Similar to the proof in Theorem~\ref{Th:prob_nonopt_scale}. For details, please refer to the supplementary  Appendix. 
\end{IEEEproof}

\begin{figure}[hbt!]
	\centering
	\includegraphics[width=0.725\textwidth]{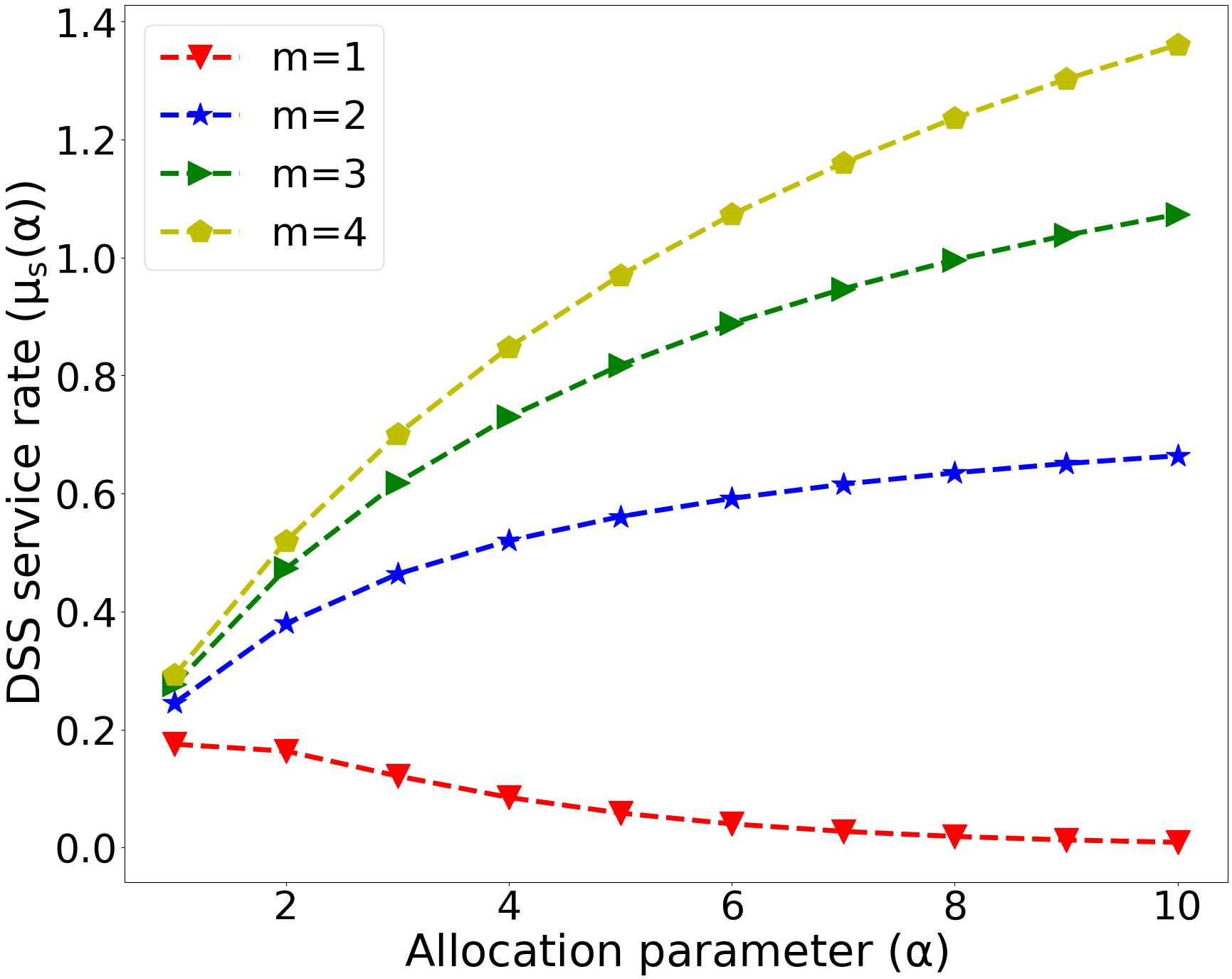}~\\
	\includegraphics[width=0.725                \textwidth]{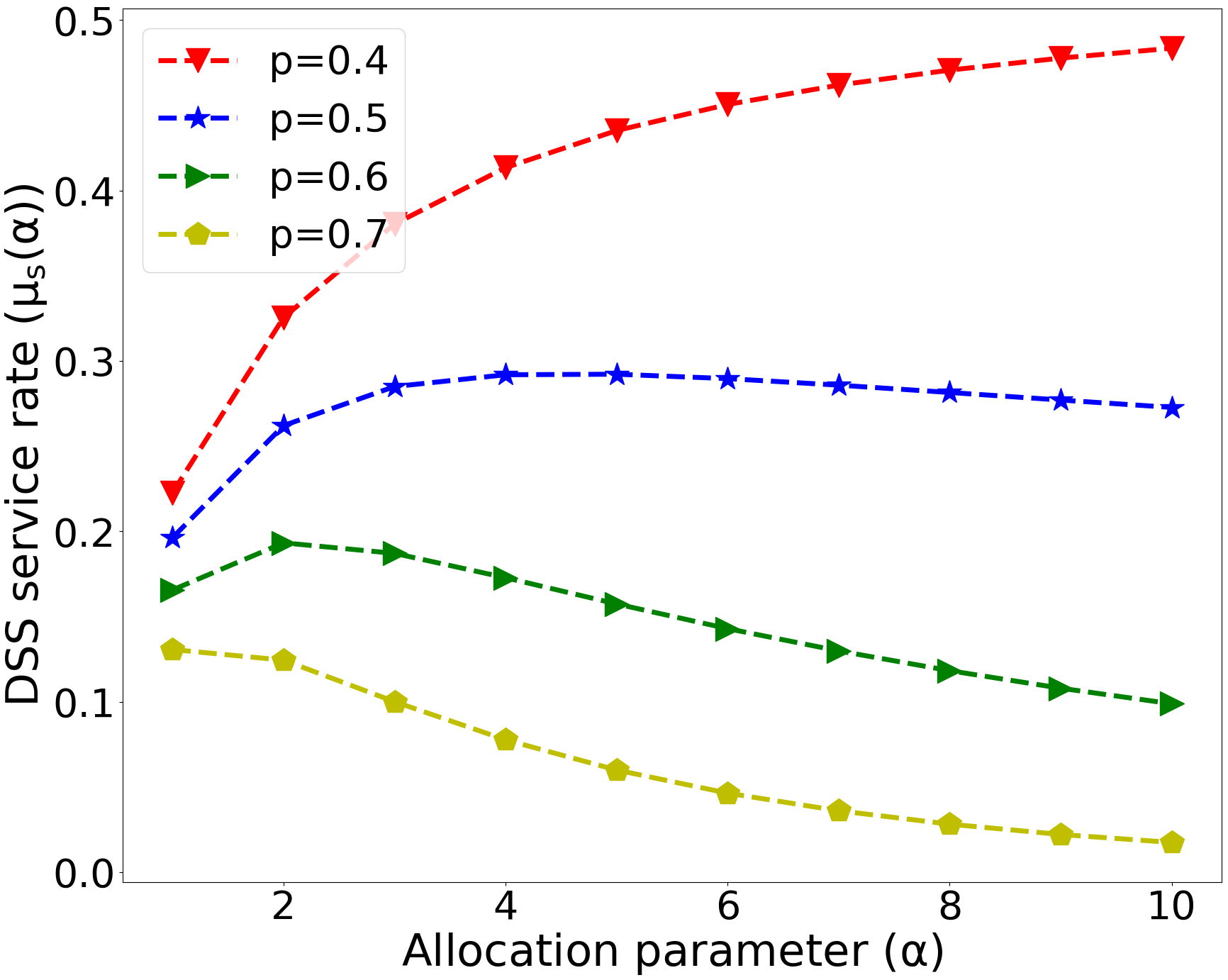}~
	\caption{ Service rate $\mu_{s}(\alpha)$ for the probabilistic access with  scaled exponential service as a function of the allocation parameter $\alpha$ (cf.~\eqref{Eq:prob_shift}).  The number of storage nodes is $N\ge m\alpha$, and the service time follows $\sexpD(3,1)$. (upper) $\mu_{s}(\alpha)$ vs. $\alpha$ for the access failure probability $p=0.3$ and four values of $m$. (lower) $\mu_{s}(\alpha)$ vs. $\alpha$ for $m=2$ and four values of $p$. Given $p$ (or $m$), the optimal allocation changes from the minimal spreading allocation to the maximal spreading allocation as $m$ increases or $p$ decreases.}
	\label{fig:large_shift_prob_mp}
\end{figure}

From Theorems~\ref{Th:prob_opt_shift} and \ref{Th:prob_nonopt_shift}, we see that whether the minimal spreading is definitely optimal or definitely not optimal depends on the probability of failed access $p$. 
The derived bounds on $p$ that guarantee the optimality or non-optimality of minimal spreading are not tight (cf.\ Remark~\ref{rem:tight}). 
However, they help us develop an insight about $\alpha$ that maximizes $\mu_s(\alpha)$, which is stated in Conjecture \ref{hyp:prob1}.

\subsubsection{Numerical Analysis}

In Fig.~\ref{fig:large_shift_prob_mp}, we evaluate the expression for $\mu_{s}(\alpha)$ given in \eqref{Eq:prob_shift} to see how the DSS service rate changes with $\alpha$. We consider a system with $N\ge m\alpha$ storage nodes and the service time follows a shifted exponential distribution with $\Delta=3$ and $\mu=1$. 
Using Theorems~\ref{Th:prob_opt_shift} and \ref{Th:prob_nonopt_shift}, we can easily calculate the optimality and non-optimality conditions. For example, when $m=2$, the minimal spreading allocation is optimal when $p>0.88$ and is non-optimal when $p< 0.08$. These conditions provide only limited knowledge of the optimal allocation and further insight on the optimal allocation can be found from Fig.~\ref{fig:large_shift_prob_mp}.
The upper graph shows $\mu_{s}(\alpha)$ vs. $\alpha$ for four different redundancy levels $m\in\{1, 2, 3, 4\}$, and the failed access probability $p=0.3$. The lower graph shows $\mu_{s}(\alpha)$ vs. $\alpha$ for four different failed access probabilities $p\in\{0.4, 0.5, 0.6, 0.7\}$, and the redundancy level $m=2$. 
In the upper subfigure, when $m=1$, the minimal spreading allocation is optimal. When $m\ge2$, the maximal spreading allocation is optimal. In the lower subfigure, when $p=0.4$, the maximal spreading allocation is optimal, while for $p=0.7$, the minimal spreading allocation is optimal. For the other two cases, an allocation with $\alpha\ge2$ is optimal. As can be seen, the optimal allocation changes with the redundancy level $m$ and the access failure probability $p$.

\begin{figure}[hbt!]
\centering
\includegraphics[width=0.725\textwidth]{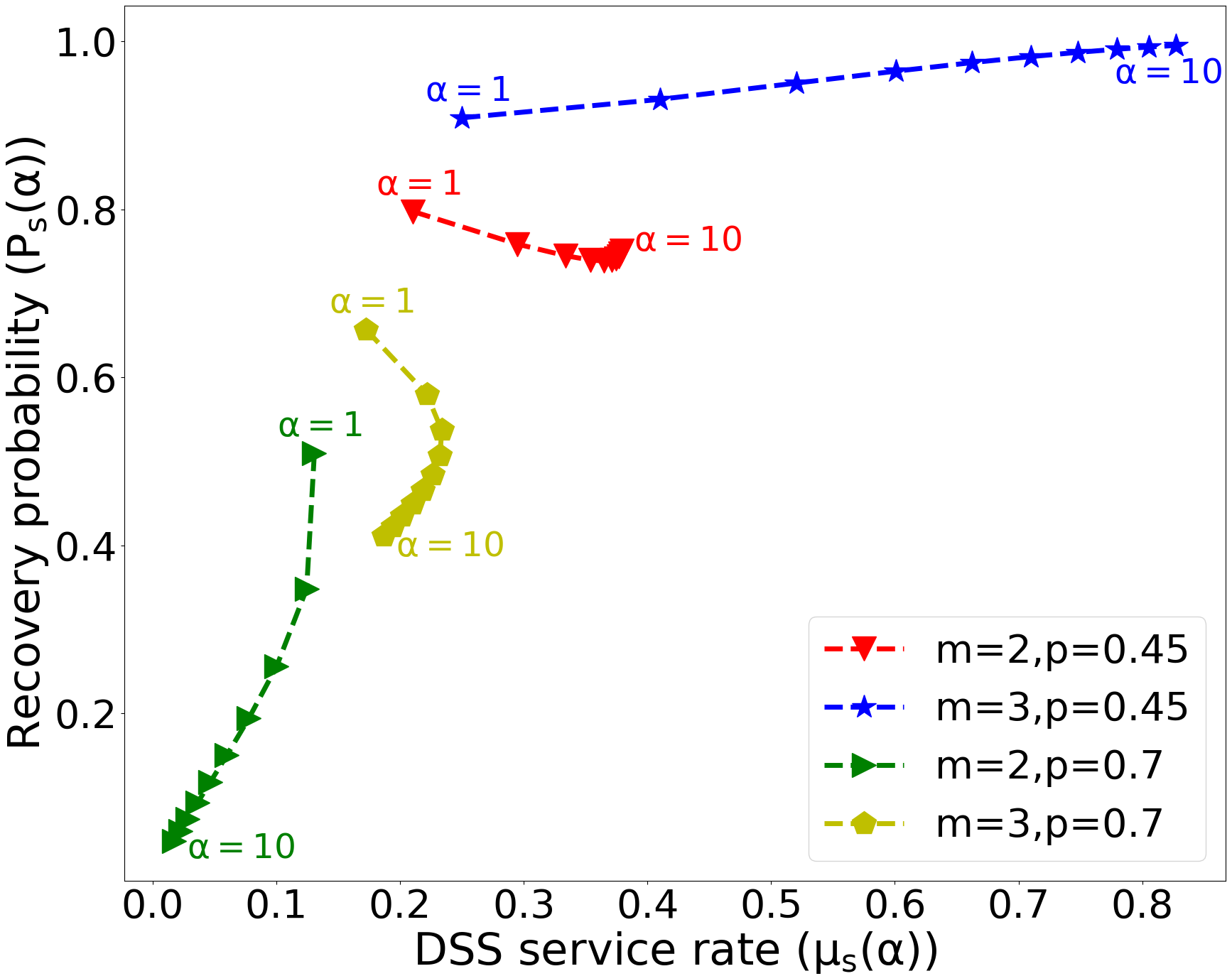}
\caption{ Successful recovery probability $P_s(\alpha)$ vs. the DSS service rate $\mu_{s}(\alpha)$ for the probabilistic access model as a function of $\alpha$ for different values of $m$ and $r$ (cf.~\eqref{eq:prob_prob} and \eqref{Eq:prob_shift}). The number of storage nodes is $N\ge m\alpha$, and the service time follows $\sexpD(3,1)$. The optimal allocation is determined by both $m$ and $r$. 
	\label{fig:large_shift_prob_up}}
\end{figure}
In Fig.~\ref{fig:large_shift_prob_up}, 
we analyze $P_s(\alpha)$ vs.\ $\mu_{s}(\alpha)$ as $\alpha$ increases from $1$ to $10$. We consider a system with $N\ge m\alpha$ nodes and two values for each parameter: $m\in\{2, 3\}$ and $p\in\{0.45, 0.7\}$. Observe that when $m$ is sufficiently small and $p$ is sufficiently large, the minimal spreading maximizes both $P_s(\alpha)$ and $\mu_{s}(\alpha)$. On the other hand, for sufficiently large $m$ and sufficiently small $p$, the maximal spreading maximizes both performance metrics. For other cases, there is no optimal $\alpha$ that maximizes both $P_s(\alpha)$ and $\mu_{s}(\alpha)$ at the same time. For example, when $m=3$ and $p=0.7$, $P_s(\alpha)$ reaches its maximum at $\alpha=1$, and $\mu_{s}(\alpha)$ reaches its maximum at $\alpha=10$.

\section{Conclusions and Future Directions}
\label{sec:con}
We considered service rates in distributed storage systems, and focused on two access models (fixed-size and probabilistic access) and two download service models (small and large file). Under the fixed-size access model, a user can access a random fixed-size subset of nodes; under probabilistic access, a user can access each node with a fixed probability. In the small file download model, the randomness associated with the file is negligible; in the large file download model, the randomness is associated with both the file size and inherent system’s operations. The primary performance metric of interest is the service rate of the system. Since redundancy for each file is fixed, the allocation of redundancy is essential for improving the system’s performance. The general allocation problem is hard to solve. We adopted the common model of quasi-uniform allocation, where coded content is uniformly spread among a subset of storage nodes. Thus the subset size completely specifies the allocation.

Minimal spreading concentrates coded chunks to a minimum-size subset. Maximal spreading allocates coded chunks to each node. For the small file model, the minimal spreading is always optimal. For the large file model, that is not the case. It is not easy to find the optimal allocation. We found the conditions under which the minimal spreading allocation is optimal. We considered scaled exponential and shifted exponential service times. Our numerical results showed that the optimal allocation under these two service models depends on the redundancy level, the number of accessed nodes, and the probability of failed access. As a general rule, one should spread the data blocks to more nodes when the redundancy level is high, the number of accessed nodes is large, or the probability of failed access is small. 
This work sets the stage for many problems of interest to be studied in the future. We briefly describe three directions of immediate interest.
\subsubsection{Optimal allocations for non-MDS codes} Our system model and analysis approach are not limited to MDS codes.
If an $[n, k]$ code with minimum distance $d$ is used, then the file can be recovered when any $\ell = n - (d - 1)$ out of $n$ blocks are downloaded. 
The Singleton bound imposes the constraint $\ell \ge k$, where the equality holds for MDS codes. When $\ell >k$, the successful recovery condition in \eqref{eq:RecCond} becomes $\sum_{i \in \mathcal{A}} s_i \ge \ell$, which means that more storage nodes must be accessed to recover the file. Thus, the exact optimal allocation values we derived may change.
\subsubsection{Optimal allocations for other service models and performance metrics} We analyzed the most common service time and scaling models. Other distributions, e.g., heavy tail Pareto and Weibull, are also of interest. For large files, it is often appropriate to model download time as the sum of the i.i.d.\ chunk download times (see \cite{peng2020diversity, peng2020diversityTIT} for additive service time scaling models). Besides, some other performance metrics, e.g., the expected download time, are also important. 
\subsubsection{Minimal spreading optimality conditions} 
We found regions of system parameters within which the minimal spreading is optimal. We expect these regions to go beyond the bounds we derived.  Finding tighter bounds would help us to better decide on when to use the minimal spreading allocation.

\appendices
\section{Proof of Theorem 10}
\begin{IEEEproof}
The proof approach is similar to the one for Theorem~$4$. 
\begin{align*}
    &\mu_s(\alpha)<\frac{\mu}{(\Delta\mu+\alpha)\binom{N}{r}}\sum^{\min(r,m\alpha)}_{\varphi=\alpha}\varphi \binom{m\alpha}{\varphi}\binom{N-m\alpha}{r-\varphi}\\
    &=\frac{\mu m\alpha\sum^{\min(r,m\alpha)}_{\varphi=\alpha}(\prod^{\alpha-2}_{i=0}\frac{m\alpha-1-i}{\varphi-1-i}) \binom{m\alpha-\alpha}{\varphi-\alpha}\binom{N-m\alpha}{r-\varphi}}{(\Delta\mu+\alpha)\binom{N}{r}}\\
    &<\frac{\mu m\alpha\sum^{\min(r,m\alpha)}_{\varphi=\alpha}(\prod^{\alpha-2}_{i=0}\frac{m\alpha-1-i}{\alpha-1-i}) \binom{m\alpha-\alpha}{\varphi-\alpha}\binom{N-m\alpha}{r-\varphi}}{(\Delta\mu+\alpha)\binom{N}{r}}\\
    &=\frac{\mu m\alpha\binom{m\alpha-1}{\alpha-1} \binom{N-\alpha}{r-\alpha}}{(\Delta\mu+\alpha)\binom{N}{r}}
\end{align*}
Similarly, we have $\mu_s(1|\varphi)\ge\frac{\mu\varphi}{\Delta\mu m+1}$, and consequently { $\mu_s(1)\ge\frac{\mu}{(\Delta\mu m+1)\binom{N}{r}}\sum^{m}_{\varphi=1}\varphi\binom{m}{\varphi}\cdot \\ \binom{N-m}{r-\varphi}=\frac{\mu m\binom{N-1}{r-1}}{(\Delta\mu m+1)\binom{N}{r}}$.} 
Now, to satisfy $\mu_s(\alpha)\le\mu_s(1)$, we have
\begin{equation}
\begin{split}
   &\frac{\mu m\alpha\binom{m\alpha-1}{\alpha-1}\binom{N-\alpha}{r-\alpha} }{(\Delta\mu+\alpha)\binom{N}{r}}\le\frac{\mu m\binom{N-1}{r-1}}{(\Delta\mu m+1)\binom{N}{r}}\\
   \Leftrightarrow &~\frac{\alpha(\Delta\mu m+1)\binom{m\alpha-1}{\alpha-1}}{\Delta\mu+\alpha}\le\prod^{\alpha-2}_{i=0}\frac{N-1-i}{r-1-i}.
\end{split}
\label{eq:shift-fix}
\end{equation}
As $\frac{N-1-i}{r-1-i}<\frac{N-2-i}{r-2-i}$ for $N>r$, it can be shown that $\prod^{\alpha-2}_{i=0}\frac{N-1-i}{r-1-i}>(\frac{N-1}{r-1})^{\alpha-1}$, and as a result, inequality \eqref{eq:shift-fix}  holds when
\begin{equation}
\begin{split}
    &\frac{\alpha(\Delta\mu m+1)\binom{m\alpha-1}{\alpha-1}}{\Delta\mu+\alpha}\le(\frac{N-1}{r-1})^{\alpha-1}\\
    \Leftrightarrow &~ r\le 1+\sqrt[\alpha-1]{\frac{\Delta\mu+\alpha}{\alpha(\Delta\mu m+1)\binom{m\alpha-1}{\alpha-1}}}(N-1).
\end{split}
\label{eq:shift-fix2}
\end{equation}
\noindent If the inequality \eqref{eq:shift-fix2} holds for all $2\le \alpha \le r$, $\mu_s(1)$ is optimal.
\end{IEEEproof}

\section{Proof of Theorem 11}

\begin{IEEEproof}
The proof approach is similar to the one for Theorem~$5$. We know $\mu_s(\alpha|\varphi)>\frac{\mu\alpha(\varphi-\alpha+1)}{\Delta\mu(m\alpha-\alpha+1)+\alpha^2}$ when $\alpha\ge2$.  Define $A=\frac{\alpha}{\Delta\mu(m\alpha-\alpha+1)+\alpha^2}$, then
\begin{align*}
    \mu_s(\alpha)&>\frac{\mu A}{\binom{N}{r}}\sum^{\min(r,m\alpha)}_{\varphi=\alpha}(\varphi-\alpha+1) \binom{m\alpha}{\varphi}\binom{N-m\alpha}{r-\varphi}\\
    &>\frac{\mu A}{\binom{N}{r}}\sum^{\min(r,m\alpha)}_{\varphi=\alpha}(\varphi-\alpha+1)  \binom{m\alpha-\alpha+1}{\varphi-\alpha+1}\binom{N-m\alpha}{r-\varphi}\\
    &=\frac{\mu A(m\alpha-\alpha+1)\binom{N-\alpha}{r-\alpha}}{\binom{N}{r}}.
\end{align*}
Similarly, we have $\mu_s(1|\varphi)\le\frac{\mu\varphi}{\Delta\mu +1}$, and consequently { $\mu_s(1)\le\frac{\mu}{(\Delta\mu +1)\binom{N}{r}}\sum^{m}_{\varphi=1}\varphi\binom{m}{\varphi}\binom{N-m}{r-\varphi}\\=\frac{\mu m\binom{N-1}{r-1}}{(\Delta\mu +1)\binom{N}{r}}$}.
Now, to satisfy $\mu_s(\alpha)\ge\mu_s(1)$, we need
\begin{equation}
\begin{split}
   &\frac{\mu A(m\alpha-\alpha+1)\binom{N-\alpha}{r-\alpha}}{\binom{N}{r}}\ge\frac{\mu m\binom{N-1}{r-1}}{(\Delta\mu +1)\binom{N}{r}}\\
  \Leftrightarrow &~ \frac{A(\Delta\mu+1)(m\alpha-\alpha +1)}{m}\ge\prod^{\alpha-2}_{i=0}\frac{N-1-i}{r-1-i}.
\end{split}
\label{eq:shift-fix3}
\end{equation}
As $\frac{N-1-i}{r-1-i}<\frac{N-2-i}{r-2-i}$ for $N>r$, we have
$\prod^{\alpha-2}_{i=0}\frac{N-1-i}{r-1-i}<(\frac{N-\alpha+1}{r-\alpha+1})^{\alpha-1}$. Inequality \eqref{eq:shift-fix3} is true when
\begin{equation}
\begin{split}
    &\frac{A(\Delta\mu+1)(m\alpha-\alpha +1)}{m}\ge(\frac{N-\alpha+1}{r-\alpha+1})^{\alpha-1}\\
    &\Leftrightarrow r\ge\sqrt[\alpha-1]{\frac{\Delta\mu m(m\alpha-\alpha+1)+m\alpha^2 }{\alpha(\Delta\mu+1)(m\alpha-\alpha+1)}}(N-\alpha+1)+\alpha-1.
\end{split}
\label{eq:shift-fix4}
\end{equation}
\noindent Therefore, if the inequality \eqref{eq:shift-fix4} holds for any one of $\alpha \in [2,r]$, $\mu_s(1)$ is not optimal.
\end{IEEEproof}

\section{Proof of Theorem 12}
\begin{IEEEproof}
The proof approach is similar to the one for Theorem~$7$. 
\begin{align*}
    &\mu_s(\alpha)<\frac{\mu}{\Delta\mu+\alpha}\sum^{m\alpha}_{\varphi=\alpha}\varphi\binom{m\alpha}{\varphi}(1-p)^{\varphi}p^{m\alpha-\varphi}\\
    &<\frac{\mu m\alpha\binom{m\alpha-1}{\alpha-1}(1-p)^{\alpha}}{\Delta\mu+\alpha}\sum^{m\alpha-\alpha}_{\varphi=0}\binom{m\alpha-\alpha}{\varphi}(1-p)^{\varphi}p^{m\alpha-\alpha-\varphi}\\
    &=\frac{\mu m\alpha\binom{m\alpha-1}{\alpha-1}(1-p)^{\alpha}}{\Delta\mu+\alpha}.
\end{align*}
Similarly, we have $\mu_s(1|\varphi)\ge\frac{\mu\varphi}{\Delta\mu m+1}$, and consequently { $\mu_s(1)>\frac{\mu}{\Delta\mu m+1}\sum^{m}_{\varphi=1}\varphi\binom{m}{\varphi}(1-p)^{\varphi}p^{m-\varphi}=\frac{\mu m(1-p)}{\Delta\mu m+1}$}.
Now, to satisfy $\mu_s(\alpha)\le\mu_s(1)$, we need
\begin{equation}
\begin{split}
   &\frac{\mu m\alpha\binom{m\alpha-1}{\alpha-1}(1-p)^{\alpha}}{\Delta\mu+\alpha}\le\frac{\mu m(1-p)}{\Delta\mu m+1}\\
   \Leftrightarrow &~ p\ge1-\sqrt[\alpha-1]{\frac{\Delta\mu+\alpha}{\alpha(\Delta\mu m+1)\binom{m\alpha-1}{\alpha-1}}}.
\end{split}
\label{eq:prob_shift_m}
\end{equation}
\noindent Therefore, if the inequality \eqref{eq:prob_shift_m} holds for all $\alpha \ge 2$, $\mu_s(1)$ is optimal.
\end{IEEEproof}

\section{Proof of Theorem 13}

\begin{IEEEproof}
The proof approach is similar to the one for Theorem~$8$. $\mu_s(\alpha|\varphi)>\frac{\mu\alpha(\varphi-\alpha+1)}{\Delta\mu(m\alpha-\alpha+1)+\alpha^2}$ when $\alpha\ge2$. Define $A=\frac{\alpha}{\Delta\mu(m\alpha-\alpha+1)+\alpha^2}$, then
\begin{align*}
    \mu_s(\alpha)&>\mu A\sum^{m\alpha}_{\varphi=\alpha}(\varphi-\alpha+1)\binom{m\alpha}{\varphi}(1-p)^{\varphi}p^{m\alpha-\varphi}\\
    &>\mu A\sum^{m\alpha}_{\varphi=\alpha}(\varphi-\alpha+1)\binom{m\alpha-\alpha+1}{\varphi-\alpha+1}(1-p)^{\varphi}p^{m\alpha-\varphi}\\
    &=\mu A(m\alpha-\alpha+1)(1-p)^{\alpha}.
\end{align*}
Similarly, we have $\mu_s(1|\varphi)\le\frac{\mu\varphi}{\Delta\mu +1}$, and consequently
{$\mu_s(1)
    <\frac{\mu}{\Delta\mu +1}\sum^{m}_{\varphi=1}\varphi\binom{m}{\varphi}(1-p)^{\varphi}p^{m-\varphi}
    =\frac{\mu m(1-p)}{\Delta\mu +1}$}.
Now, to satisfy $\mu_s(\alpha)\ge\mu_s(1)$, we need
\begin{equation}
\begin{split}
   &\mu A(m\alpha-\alpha+1)(1-p)^{\alpha}\ge\frac{\mu m(1-p)}{\Delta\mu +1}\\
   \Leftrightarrow &~ p\le1-\sqrt[\alpha-1]{\frac{m(\Delta\mu(m\alpha-\alpha+1)+\alpha^2)}{\alpha(\Delta\mu +1)(m\alpha-\alpha+1)}}.
\end{split}
\label{eq:shift_prob_m2}
\end{equation}
\noindent Therefore, if the inequality \eqref{eq:shift_prob_m2} holds for $\alpha \ge 2$, $\mu_s(1)$ is not optimal.
\end{IEEEproof}

\bibliographystyle{IEEEtran}
\bibliography{distributed}

\begin{thebibliography}{10}
\providecommand{\url}[1]{#1}
\csname url@samestyle\endcsname
\providecommand{\newblock}{\relax}
\providecommand{\bibinfo}[2]{#2}
\providecommand{\BIBentrySTDinterwordspacing}{\spaceskip=0pt\relax}
\providecommand{\BIBentryALTinterwordstretchfactor}{4}
\providecommand{\BIBentryALTinterwordspacing}{\spaceskip=\fontdimen2\font plus
\BIBentryALTinterwordstretchfactor\fontdimen3\font minus
  \fontdimen4\font\relax}
\providecommand{\BIBforeignlanguage}[2]{{%
\expandafter\ifx\csname l@#1\endcsname\relax
\typeout{** WARNING: IEEEtran.bst: No hyphenation pattern has been}%
\typeout{** loaded for the language `#1'. Using the pattern for}%
\typeout{** the default language instead.}%
\else
\language=\csname l@#1\endcsname
\fi
#2}}
\providecommand{\BIBdecl}{\relax}
\BIBdecl

\bibitem{noori2016storage}
M.~Noori, E.~Soljanin, and M.~Ardakani, ``On storage allocation for maximum
  service rate in distributed storage systems,'' in \emph{2016 IEEE Internat.\
  Symp.\ on Information Theory (ISIT)}, 2016, pp. 240--244.

\bibitem{peng2018distributed}
P.~Peng and E.~Soljanin, ``On distributed storage allocations of large files
  for maximum service rate,'' in \emph{2018 56th Annual Allerton Conference on
  Communication, Control, and Computing (Allerton)}, 2018, pp. 784--791.

\bibitem{joshi2012coding}
G.~Joshi, Y.~Liu, and E.~Soljanin, ``Coding for fast content download,'' in
  \emph{2012 50th Annual Allerton Conference on Communication, Control, and
  Computing (Allerton)}.\hskip 1em plus 0.5em minus 0.4em\relax IEEE, 2012, pp.
  326--333.

\bibitem{kadhe15availability}
S.~Kadhe, E.~Soljanin, and A.~Sprintson, ``Analyzing the download time of
  availability codes,'' in \emph{2015 IEEE Internat.\ Symp.\ on Information
  Theory (ISIT)}.\hskip 1em plus 0.5em minus 0.4em\relax IEEE, 2015, pp.
  1467--1471.

\bibitem{aktas2019straggler}
M.~Aktas and E.~Soljanin, ``Straggler mitigation at scale,'' \emph{IEEE/ACM
  Trans.\ Networking}, vol.~27, no.~06, pp. 2266--2279, Nov. 2019.

\bibitem{aktas2021download}
M.~F. Akta{\c{s}}, S.~Kadhe, E.~Soljanin, and A.~Sprintson, ``Download time
  analysis for distributed storage codes with locality and availability,''
  \emph{IEEE Transactions on Communications}, 2021.

\bibitem{Edge:YadgarKAS19}
G.~Yadgar, O.~Kolosov, M.~F. Aktas, and E.~Soljanin, ``Modeling the edge:
  Peer-to-peer reincarnated,'' in \emph{2nd {USENIX} Workshop on Hot Topics in
  Edge Computing, HotEdge 2019, Renton, WA, USA, July 9, 2019}.\hskip 1em plus
  0.5em minus 0.4em\relax {USENIX} Association, 2019.

\bibitem{leong2011distributed}
D.~Leong, A.~G. Dimakis, and T.~Ho, ``Distributed storage allocations for
  optimal delay,'' in \emph{2011 IEEE Internat.\ Symp.\ on Information Theory
  Proceedings}.\hskip 1em plus 0.5em minus 0.4em\relax IEEE, 2011, pp.
  1447--1451.

\bibitem{peng2020diversity}
P.~Peng, E.~Soljanin, and P.~Whiting, ``Diversity vs. parallelism in
  distributed computing with redundancy,'' in \emph{2020 IEEE Internat.\ Symp.\
  on Information Theory (ISIT)}.\hskip 1em plus 0.5em minus 0.4em\relax IEEE,
  2020, pp. 257--262.

\bibitem{peng2020diversityTIT}
------, ``Diversity/parallelism trade-off in distributed systems with
  redundancy,'' \emph{arXiv preprint arXiv:2010.02147}, 2020.

\bibitem{leong2012distributed}
D.~Leong, A.~G. Dimakis, and T.~Ho, ``Distributed storage allocations,''
  \emph{IEEE Trans.\ Information Theory}, vol.~58, no.~7, pp. 4733--4752, 2012.

\bibitem{Sardari_Allocation_2010}
M.~Sardari, R.~Restrepo, F.~Fekri, and E.~Soljanin, ``Memory allocation in
  distributed storage networks,'' in \emph{2010 IEEE Internat.\ Symp.\ on
  Information Theory}.\hskip 1em plus 0.5em minus 0.4em\relax IEEE, 2010, pp.
  1958--1962.

\bibitem{hong2014asymptotic}
B.~Hong and W.~Choi, ``Asymptotic analysis of failed recovery probability in a
  distributed wireless storage system with limited sum storage capacity,'' in
  \emph{2014 IEEE Internat.\ Conf.\ on Acoustics, Speech and Signal Processing
  (ICASSP)}, 2014, pp. 6459--6463.

\bibitem{noori2015allocation}
M.~Noori and M.~Ardakani, ``Allocation for heterogeneous storage nodes,''
  \emph{IEEE Communic.\ Letters}, vol.~19, no.~12, pp. 2102--2105, 2015.

\bibitem{wu2020optimal}
F.~Wu, Y.~Wang, C.~Zhang, W.~Fan, and Y.~Liu, ``Optimal storage allocation for
  delay sensitivity data in electric vehicle network,'' \emph{IEEE Transactions
  on Intelligent Transportation Systems}, vol.~22, no.~1, pp. 579--591, 2020.

\bibitem{matching:AlonFHRRS12}
N.~Alon, P.~Frankl, H.~Huang, V.~R{\"o}dl, A.~Ruci{\'n}ski, and B.~Sudakov,
  ``Large matchings in uniform hypergraphs and the conjectures of {E}rd{\H{o}}s
  and {S}amuels,'' \emph{Journal of Combinatorial Theory, Series A}, vol. 119,
  no.~6, pp. 1200--1215, 2012.

\bibitem{service:aktas2020service}
M.~Aktas, G.~Joshi, S.~Kadhe, F.~Kazemi, and E.~Soljanin, ``Service rate
  region: A new aspect of coded distributed system design,'' \emph{arXiv
  preprint arXiv:2009.01598}, 2020.

\bibitem{service:KazemiKSS21g}
F.~Kazemi, S.~Kurz, E.~Soljanin, and A.~Sprintson, ``Efficient storage schemes
  for desired service rate regions,'' \emph{2021 IEEE Inform.\ Th.\ Workshop
  (ITW)}, Apr. 2021.

\bibitem{service:KazemiKSS20g}
F.~Kazemi, S.~Kurz, and E.~Soljanin, ``A geometric view of the service rates of
  codes problem and its application to the service rate of the first order
  reed-muller code,'' \emph{2020 IEEE Internat.\ Symp.\ on Information Theory
  (ISIT)}, June 2020.

\bibitem{service:KazemiKSS20c}
F.~Kazemi, E.~Karimi, E.~Soljanin, and A.~Sprintson, ``A combinatorial view of
  the service rates of codes problem, its equivalence to fractional matching
  and its connection with batch codes,'' \emph{2020 IEEE Internat.\ Symp.\ on
  Information Theory (ISIT)}, June 2020.

\bibitem{service:AndersonJJ18}
S.~E. Anderson, A.~Johnston, G.~Joshi, G.~L. Matthews, C.~Mayer, and
  E.~Soljanin, ``Service capacity region of content access from erasure coded
  storage,'' \emph{IEEE Inform.\ Th.\ Workshop (ITW)}, Nov. 2018.

\bibitem{service:AktasAJ17}
M.~Akta{\c{s}}, S.~E. Anderson, A.~Johnston, G.~Joshi, S.~Kadhe, G.~L.
  Matthews, C.~Mayer, and E.~Soljanin, ``On the service capacity region of
  accessing erasure coded content,'' in \emph{2017 55th Annual Allerton Conf.\
  Commun., Control, and Comput.\ (Allerton)}, 2017, pp. 17--24.

\bibitem{service:RaaijmakersB20}
Y.~Raaijmakers and S.~C. Borst, ``Achievable stability in redundancy systems,''
  \emph{Proc. {ACM} Meas. Anal. Comput. Syst.}, pp. 1--21, 2020.

\bibitem{arnold2008first}
B.~C. Arnold, N.~Balakrishnan, and H.~N. Nagaraja, \emph{A first course in
  order statistics}.\hskip 1em plus 0.5em minus 0.4em\relax SIAM, 2008.

\end{thebibliography}
\begin{IEEEbiography}[{\includegraphics[width=1in,height=1.25in,clip,keepaspectratio]{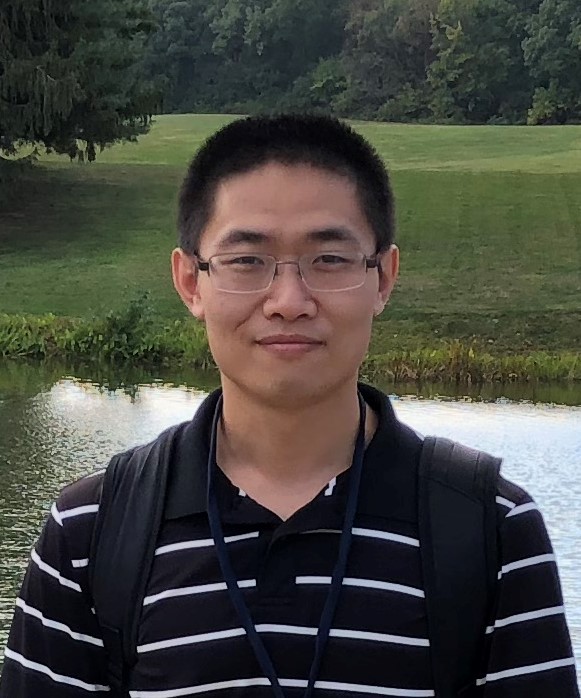}}]{Pei Peng}
Pei Peng received his B.S. degree in information engineering from South China University of Technology, Guangzhou, China, in 2011, and the M.S. degree in electronics and communication engineering from Shanghai Institute of Micro-system and Information Technology, Chinese Academy of Science, Shanghai, China, in 2014. 
Since 2015, he has been a Ph.D.\ student in the electrical and computer engineering department at Rutgers, the State University of New Jersey, USA. During his PhD.\ studies, Pei Peng has served as both research and teaching assistant and has received a 2020 ECE department teaching award. His research interests are coding and allocation in distributed computing systems, covert communications, and machine learning.

\end{IEEEbiography}

\begin{IEEEbiography}[{\includegraphics[width=1in,height=1.25in,clip,keepaspectratio]{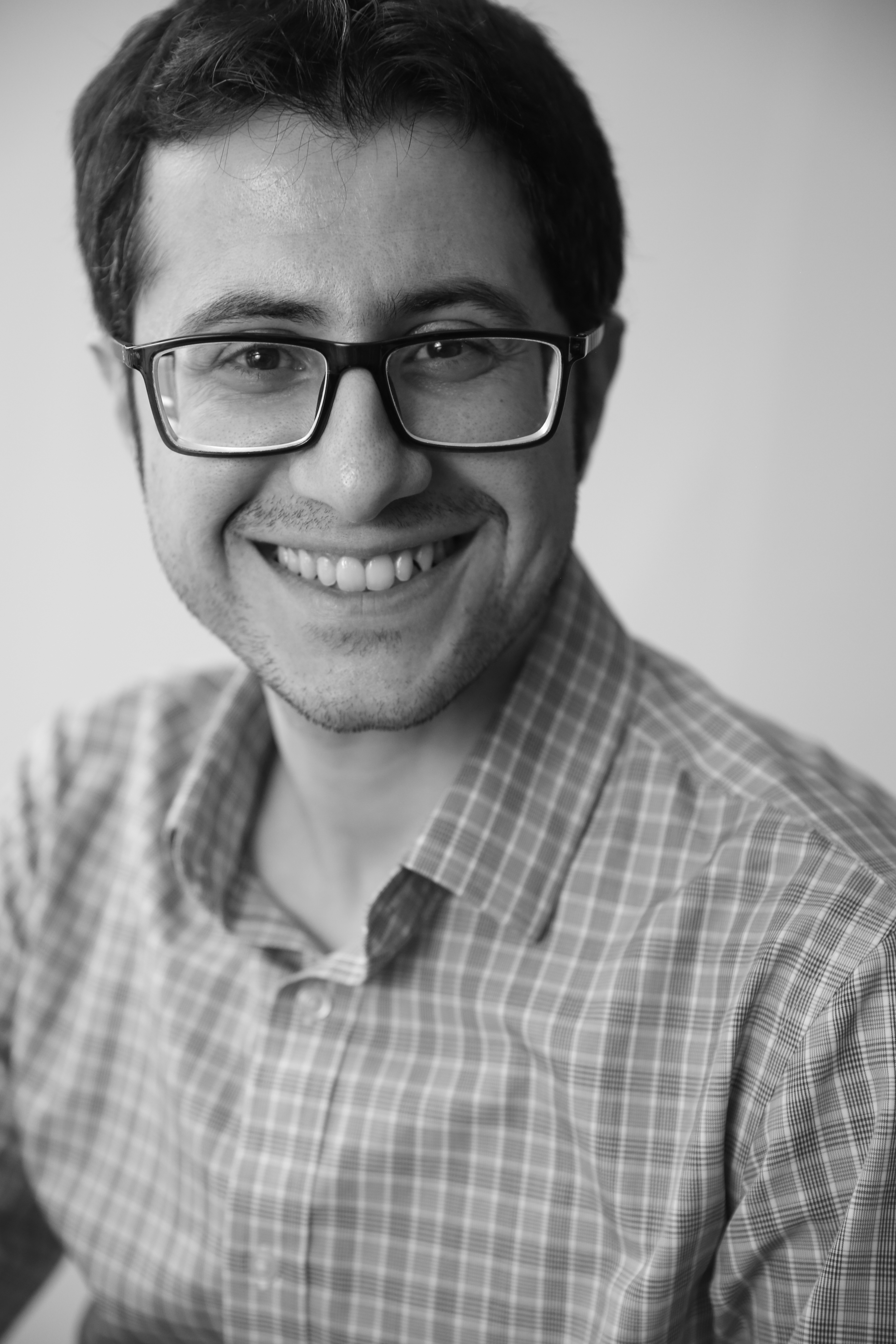}}]{Moslem Noori}
Moslem Noori (S’07–M’13) received the B.Sc. degree in electrical engineering and the B.Sc. degree in applied mathematics from the Amirkabir University of Technology, in 2005 and 2006, respectively, and the M.Sc. and Ph.D. degrees in electrical engineering from the University of Alberta, in 2008 and 2012, respectively. He held a postdoctoral position with The University of British Columbia from 2013 to 2014. He returned to the University of Alberta as an Alberta Innovates Technology Futures Post-Doctoral Fellow from 2014 to 2016. He is currently a principal scientist at 1QB Information Technologies. His research interests include distributed storage systems, wireless communications, quantum computation and communication, machine learning for medical applications, and stochastic network analysis. He has received several awards and scholarships, including the NSERC Vanier CGS, the NSERC Post-Doctoral Fellowship, and the AITF Post-Doctoral Fellowship.
\end{IEEEbiography}

\begin{IEEEbiography}[{\includegraphics[width=1in,height=1.25in,clip,keepaspectratio]{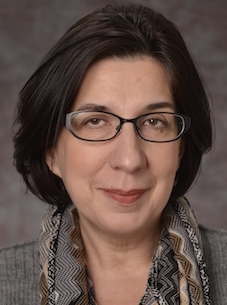}}]{Emina Soljanin}
is a professor at Rutgers University. Before moving to Rutgers in 2016, she was a (Distinguished) Member of Technical Staff for 21 years in the Mathematical Sciences Research of Bell Labs.
Her interests and expertise are wide. Over the past quarter of the century, she has participated in numerous research and business projects, as diverse as power system optimization, magnetic recording, color space quantization, hybrid ARQ, network coding, data and network security, distributed systems performance analysis, and quantum information theory. She served as an Associate Editor for Coding Techniques, for the IEEE Transactions on Information Theory, on the Information Theory Society Board of Governors, and in various roles on other journal editorial boards and conference program committees.
 Prof.~Soljanin an IEEE Fellow, an outstanding alumnus of the Texas A\&M School of Engineering, the 2011 Padovani Lecturer, a 2016/17 Distinguished Lecturer, and 2019 President of the IEEE Information Theory Society.
 \end{IEEEbiography}
\end{document}